\documentclass[12pt]{article} 
     \usepackage{amsmath}
\usepackage{mathtools}
\usepackage{units}
\usepackage[T1]{fontenc}
\usepackage{xspace}
\newcommand{\rrangle}{>\kern-1.2ex~>\xspace}
\newcommand{\llangle}{<\kern-1.2ex~<\xspace}

\usepackage{amsfonts,amsthm,amsmath,amssymb,upgreek,bm}
\usepackage{hyperref}
\usepackage[paper=letterpaper,margin=.7in]{geometry}
\usepackage{units}
\usepackage{float}
\usepackage[export]{adjustbox}
\usepackage{bbold}
\usepackage{enumerate}%
\usepackage{xcolor}
\usepackage[shortlabels]{enumitem}
\usepackage{dsfont}

\newcommand{\olsi}[1]{\,\overline{\!{#1}}}

  \usepackage{graphicx}
  \usepackage{color}
  \newlength{\abstractwidth}
 \usepackage{dsdshorthand}

 \usepackage{amsmath,amssymb}
\usepackage{graphicx}
\usepackage{subfigure}
\usepackage{caption}
\usepackage{draft}
\usepackage{lipsum}
\usepackage[utf8]{inputenc}
\usepackage{scalerel}
  \usepackage{mciteplus} 
\usepackage{hyperref}
\usepackage{cancel}
\usepackage{tikz}
\usetikzlibrary{arrows,positioning} 
\tikzset{
    >=stealth',
    punkt/.style={
           rectangle,
           rounded corners,
           draw=black, very thick,
           text width=15em,
           minimum height=2em,
           text centered},
    pil/.style={
           ->,
           thick,
           shorten <=2pt,
           shorten >=2pt,}
}

  \newcommand{\beq}{\begin{equation}}
  \newcommand{\eeq}{\end{equation}}
  
  \renewcommand{\>}{\rangle}

  \def\ba{\begin{eqnarray}}
  \def\ea{\end{eqnarray}}

 \def\simleq{\; \raise0.3ex\hbox{$<$\kern-0.75em
      \raise-1.1ex\hbox{$\sim$}}\; }
 \def\simgeq{\; \raise0.3ex\hbox{$>$\kern-0.75em
      \raise-1.1ex\hbox{$\sim$}}\; }

\begin{document}

\begin{titlepage}
  \bigskip

  \bigskip\bigskip

  \bigskip

\begin{center}
%\centerline
{\Large \bf{A violation of global symmetries from replica wormholes\\\vspace{0.5cm}and the fate of black hole remnants}}
 \bigskip
%\centerline
{\Large \bf { }} 
    \bigskip
\bigskip
\end{center}

\preprint{CALT-TH-2020-051}

  \begin{center}

 \bf {Po-Shen Hsin$^1$, Luca V. Iliesiu$^{2}$  and Zhenbin Yang$^2$  }
  \bigskip \rm
\bigskip
 
 $^1$ Walter Burke Institute for Theoretical Physics,
California Institute of Technology, Pasadena, CA 91125, USA\\
$^2$Stanford Institute for Theoretical Physics, Stanford University, Stanford, CA 94305, USA

 \rm

\bigskip
\bigskip

  \end{center}

 \bigskip\bigskip
  \begin{abstract}

 \medskip
  \noindent
  \end{abstract}
We show that the presence of replica wormholes in the Euclidean path integral of gravity leads to a non-perturbative violation of charge conservation for any global symmetry present in the low-energy description of quantum gravity. Explicitly, we compute the scattering probability between different charged states in several two-dimensional models of quantum gravity and find a non-vanishing answer. This suggests that the set of all charged states is typically over-complete, which has drastic consequences for the fate of black hole remnants that could carry a global symmetry charge.  In the holographic context, we argue that the presence of such a symmetry in the effective description of the bulk should appear on the boundary as an emergent global symmetry after ensemble averaging.
\bigskip \bigskip \bigskip

  \end{titlepage}

   \tableofcontents
\newpage
 \section{Introduction}
 \label{sec:intro}
 
A longstanding question in quantum gravity is whether 
exact global symmetries can be present \cite{Misner:1957mt, Giddings:1988cx, Kallosh:1995hi, Polchinski:2003bq, ArkaniHamed:2006dz, Banks:2010zn, Harlow:2018jwu, Harlow:2018tng, Harlow:2020bee}. 
Global symmetries provide an important guiding principle to organize universality classes of effective field theories and they are fundamental to our understanding of particle physics and of phases of matter.
However, when such global symmetries are present in quantum gravity, a problem appears when forming black holes from particles that carry an overall global symmetry charge. 
A semi-classical analysis of evaporation suggests that the Hawking radiation emitted by such black holes is thermal \cite{HAWKING1974, Hawking1975, Hawking:1982dj}. Consequently, the global symmetry charge of the black hole cannot significantly change through the evaporation process. If the black hole evaporates completely and the resulting state of Hawking radiation is neutral under the global symmetry (as suggested by the semi-classical analysis), then the evaporation process violates the conservation of the global symmetry charge, a central tenet of the global symmetries in quantum field theory.  Alternatively, the black hole might not evaporate completely but rather decay to some remnant state that retains the global symmetry charge of the black hole. Because in the process of evaporation, a large number of indistinguishable small black holes can be formed \cite{Banks:2010zn}, each with a different global symmetry charge,\footnote{For now, we assume that the global symmetry is continuous or, if discrete, has a large number of irreducible unitary representations. } such states have a much larger degeneracy than what is allowed by the ``central dogma''  \cite{Almheiri:2020cfm}, which states that from the perspective of an outside observer, black holes describe quantum systems with $S_\text{BH} = (\text{horizon area})/(4G_{\text N})$ degrees of freedom.

The primary goal of this paper is to show that if a global symmetry is present in the effective description in any theory of quantum gravity,\footnote{Here, we can consider the case where the global symmetry is present in the effective description up to arbitrarily large energy scales.} then the presence of (replica) wormholes in the Euclidean path integral of the theory leads to a non-perturbative violation of this global symmetry.\footnote{There are other works in the literature that discussed the
connection between the existence of wormhole solutions and the violation of global symmetries in quantum gravity \cite{Kallosh:1995hi, ABBOTT1989687}. However, the wormholes discussed in this paper are different in nature, {\it i.e.}~here they appear in the gravitational path integral due to the presence of multiple components of boundaries.   }  

We examine the violation of charge conservation by computing the scattering probability between states with different global symmetry charges. In a black hole background, we discover that this probability is non-zero.  
This is in contrast to the common intuition from quantum field theory: instead of living in different superselection sectors with zero overlapping probability, 
the states with different global symmetry charges in quantum gravity are non-orthogonal and could form an over-complete basis. 
Consequently, to understand whether small black holes or remnants indeed disobey the  ``central dogma'' discussed above, we determine the minimal basis of charged states that spans the space of states for such objects. 
In contrast to the previous analysis, which suggests that the dimension of this basis is equal to the number of unitary irreducible representations of the global symmetry group \cite{Banks:2010zn},\footnote{Of course, in the case of a continuous symmetry group, this number is infinite.  }  we find that the dimension is always given by $\sim \,e^{S_\text{BH}}$ after considering the contribution of connected geometries in the gravitational path integral.\footnote{
In particular, at late time of the black hole evaporation with small $S_\text{BH}$, the symmetry violation can be observed in the measurement of the overlap probability between a reasonably small number of sectors with different charges.}
The result is now consistent with the  ``central dogma'' and suggests that, in principle, black holes that carry a global symmetry charge can fully evaporate. 
 
 It is perhaps not surprising that the contribution of (replica) wormholes to the gravitational path integral drastically alters the conclusions of the semi-classical analysis for black hole evaporation. Recently, by considering the contribution of replica wormholes, \cite{Penington:2019kki, Almheiri:2019qdq, Almheiri:2020cfm} reproduced the correct behavior of the Page curve at late times, further providing a detailed map of how modes trapped in the interior of the black hole are encoded in the Hawking radiation at late times. The analysis presented in this paper is closely related to these developments, as we explain how global symmetry charges trapped inside the black hole horizon can ``escape'' through replica wormholes analogous to those considered in \cite{Penington:2019kki, Almheiri:2019qdq}.

 Even though the conservation of the global symmetry charge is violated due to wormhole contributions, one might ponder the origin of this symmetry in the effective gravitational theory, in the context of holography. For bulk theories with multiple boundaries, the contribution of wormholes to the gravitational path integral leads to the widely discussed factorization puzzle:  from the field
theory point of view the correlation functions across two boundaries seemly factorize, while from the gravity point of view they do not \cite{Maldacena:2004rf}. To obtain boundary observables consistent with the lack of factorization in the bulk, we can consider a boundary system given by an ensemble average of theories \cite{Maldacena:2004rf}. In this context, we conjecture the following relation between the bulk and the boundary:
 \begin{itemize}
     \item If each theory in the ensemble average on the boundary has some global symmetry $G$, then the effective theory in the bulk should have a \textbf{gauge} symmetry whose gauge group is also given by $G$. This is the standard case in AdS/CFT \cite{Witten:1998qj}.
     
    \item If the ensemble average on the boundary gives rise to some emergent global symmetry $G$ (which is not a symmetry of individual Hamiltonians in the ensemble), then the effective gravitational theory in the bulk should have the same \textbf{global} symmetry $G$. On the boundary, global symmetry charge conservation is violated when considering the average of several replicas of the ensemble, while in the bulk the violation occurs because of replica wormholes. This means that if we couple this boundary symmetry to a background gauge field, the replicated system will not be invariant under the most general gauge transformations. A concrete example to have in mind for a boundary theory exhibiting such features is the Sachdev-Ye-Kitaev (SYK) model \cite{Sachdev:2015efa, kitaevTalks, Maldacena:2016hyu, Kitaev:2017awl}; the model has an emergent $O(N)$ symmetry after ensemble averaging but has no such symmetry in individual instances of the ensemble. 
 \end{itemize}
 
 This improvement of the holographic dictionary provides a new perspective on the problem of factorization. If we want to restore the factorization in the bulk with modifications of the Lagrangian of the theory or by finding some UV completion, this requires an explicit breaking of all bulk global symmetries in order to be consistent with the wormhole calculation.

The remainder of this paper is organized as follows. In section \ref{sec:review-past}, we review several known arguments that suggest the absence of global symmetries in quantum gravity. In section \ref{sec:replica-wormhole-and-global-symm}, we present the general setup for our calculation and compute the nonzero scattering amplitude between states of different charges in several toy models of gravity. Furthermore, we emphasize the difference between global and gauge symmetries in the gravitational path integral, explaining why the latter does not exhibit a violation of charge conservation, while the former does. We also discuss how global and gauge symmetries can arise in holographic theories with ensemble-averaged dual boundary theories. In section \ref{sec:state-reconst}, we show that replica wormhole predicts that any $e^{S_{\text{BH}}}$ number of semi-classical states in the black hole interior span the complete Hilbert space of black hole.
In particular, this means that the semi-classical states behind the black hole horizon are over-complete and any other excitation in the black hole interior can be reconstructed from these $e^{S_{\text{BH}}}$ states.
In Jackiw-Teitelboim (JT) gravity \cite{Teitelboim:1983ux,Jackiw:1984je}, we do an exact planar resummation to find the reconstruction map, and explicitly find the complete basis of states.
In section \ref{sec:BH-remn} we relate our findings to the problem of remnants and we speculate about their ultimate fate.
Finally, in section \ref{sec:conclusion} we summarize the main points of our paper and discuss their relation to past arguments against global symmetries in quantum gravity. 
\\

 \textbf{Note added:} During the development of this paper,  \cite{Chen:2020ojn} appeared which, using a different perspective, also discussed the violation of global symmetries in quantum gravity using replica wormholes.

  \section{Review of previous arguments}
  \label{sec:review-past}
  
\begin{figure}[h]
\begin{center}
\begin{tikzpicture}[ scale=1.0]
 \pgftext{\includegraphics[scale=0.5]{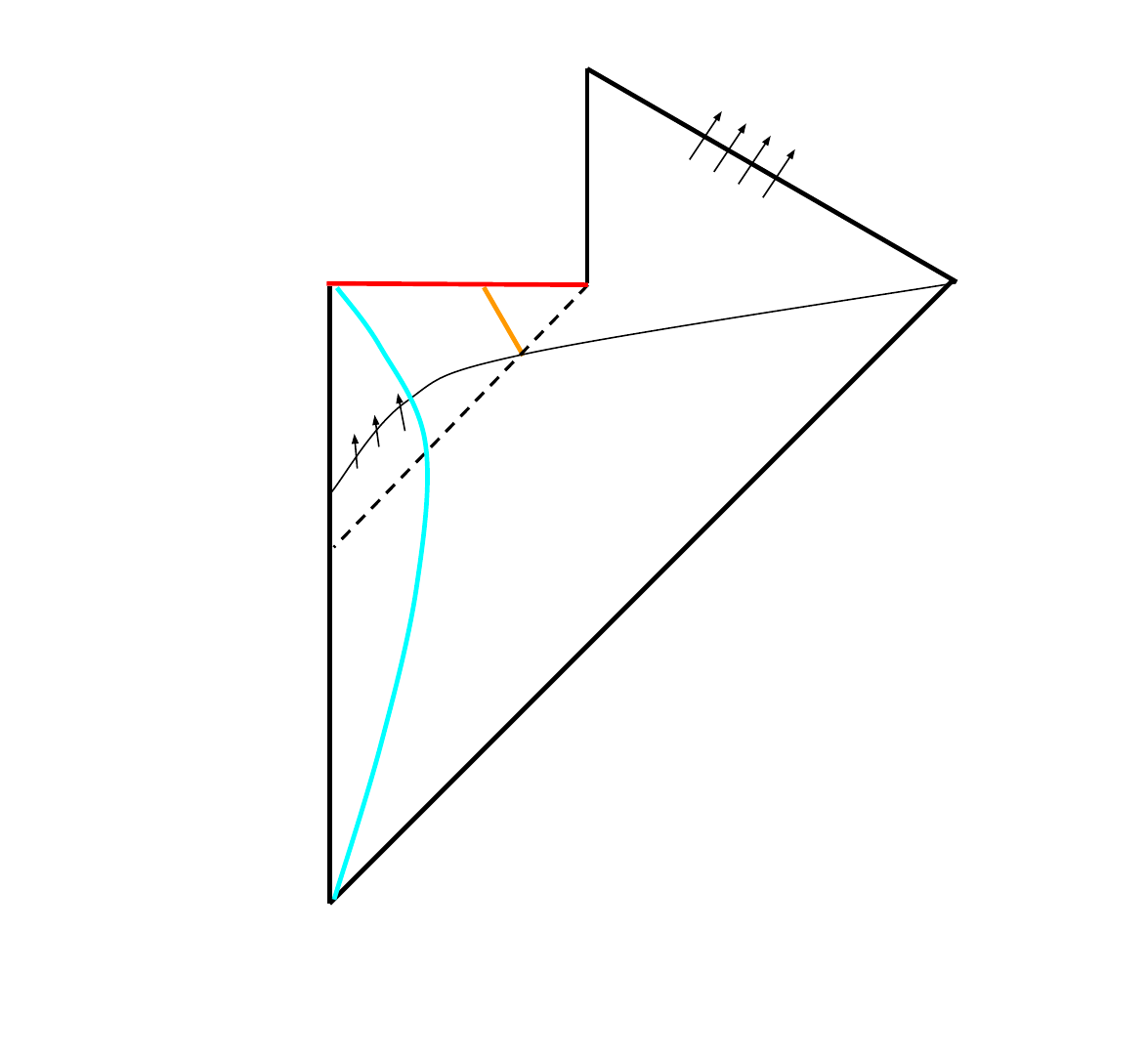}} at (0,0);
 \draw (-1.9,-3.85) node  {$\mathcal I^-$};
 \draw (-1.1,1.65) node  {$\mathcal H$};
 \draw (2.2,+3.6) node  {$\mathcal{J}^+ $};
  \draw (-2.0,-1.5) node {Charged};
   \draw (-2.0,-1.85) node {matter};
\draw (2.0,+2.5) node {Neutral};
   \draw (2.0,2.2) node {state};
\end{tikzpicture}
\caption{ The Penrose Diagram of an evaporating black hole.  At $\mathcal{I}^-$, we collide a large amount of particles, forming a representation $R$ under the global symmetry $G$, to create a black hole. The Hawking radiation at $\mathcal{J}^+$ is thermal and independent of $R$ \cite{HAWKING1974,Hawking1975}.  The original matter is stored in the interior region $\mathcal{H}$, which causes a problem when the log of dimension of the representation $R$ exceeds the Bekenstein-Hawking entropy.}
\label{fig:Penrose Diagram}
\end{center}
\end{figure}
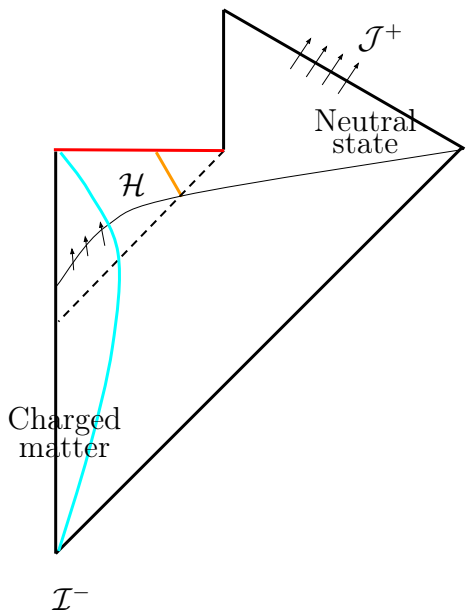

    Before expanding on our arguments regarding the violation of global symmetries due to replica wormholes, it is instructive to first review several arguments about the absence of the global symmetry in quantum gravity.  We can summarize most of these arguments through figure \ref{fig:Penrose Diagram}, that shows the Penrose diagram of an evaporating black hole. As we will discuss later in the paper, each one of the arguments reviewed below can be refined by including the contribution of connected geometries to the gravitational path integral. 
    
  \subsection{Hawking's original argument}
  \label{sec:hawking-arg}

 Considering a quantum field theory with a continuous global symmetry coupled to gravity, we can create a black hole by colliding a shell of matter field carrying an overall non-zero global charge.
In the absence of gauge fields for the symmetry ({\it i.e.} the symmetry is not gauged),
the macroscopic structure of the black hole is insensitive to the global symmetry charge due to the no-hair theorem \cite{Israel:1967wq,Israel:1967za,Carter:1971zc,Robinson:1975bv}.
This means that the Hawking radiation will be the same as the case of an ordinary Schwarzchild black hole.
In the leading approximation, only the lightest particles are produced in the evaporation, and they do not carry any of the original global symmetry charges and at the end of the evaporation process, assuming there are no remnants that carry the global symmetry charge, we have an almost thermal distribution of radiation.
This means that the black hole evaporation process violates the charge conservation of the symmetry, and therefore global symmetry must be violated in quantum gravity \cite{Hawking:1982dj}.
The argument can be formulated quantitatively as a
scattering process, by defining
a dollar matrix $\$$ that represents the transition amplitude between the initial and final density matrix \cite{Hawking:1982dj}:
\be
\$ _{m m'; n n'} \rho^{in}_{n n'}=\rho^{out}_{m m'},
\ee
which for a unitary process can be factorized as a product of S-matrix elements: 
\be\label{eqn:dollar matrix}
\$_{m m';n n'}=S_{m n} S^*_{m' n'}.
\ee
Here we used notation $m,n$ to label the coordinates of the Hilbert space.
For a single non-unitary evolution, the dollar matrix cannot be factorized.
 Then, the breaking of a global symmetry $G$ means that the dollar matrix does not commute with the symmetry transformation:
 \be
\$_{m m'; n n'} G_{n l}G^*_{n'l'}\neq G_{m n} G^*_{m' n'}\$_{n n'; l l'} ~.
 \ee
 The above argument assumes that the black hole fully evaporates and no remnant is present. This, of course might be incorrect since the semi-classical computation of the rate of Hawking radiation can fail once the black hole mass becomes sufficiently small,  say  $M = X M_{\text pl}$ for some number $X$.  As we will review below, the absence of global symmetries in quantum gravity can be motivated even when 
 assuming the presence of remnants.  

  \subsection{The remnant argument}
  \label{sec:banks-seiberg-arg}
  
 While Hawking's original argument is physically intuitive, it relies on the assumption that nothing dramatic happens at the end point of the evaporation process where the semi-classical calculation may break down.
 The validity of Hawking's calculation can be estimated from the change of the black hole mass due to radiating one thermal quanta.
 For example, the temperature of a mass $M$ Schwarzchild black hole is of order $1/ (G_{\text N}M)$, which means that the semi-classical picture breaks down when $M\sim M_{\text pl}$.
 
 Banks and Seiberg, on the other hand, provide an additional argument against global symmetries \cite{Banks:2010zn}.
Just like the setup described above, they imagine the initial matter forms a representation $R$ of a large dimension under some non-Abelian group $G$.
 Under the black hole evaporation, the initial matter will remain in the interior region $\mathcal{H}$ of the black hole.
 The black hole can evaporate down to a mass $M = X M_{\text pl}$,  where $X M_{\text pl}$ is the energy scale at which semi-classical thermodynamics  breaks down and the result might be a long lived remnant. At this energy scale the entropy on its lightsheet (the non-expanding lightcone associated to its horizon) will be order $\sim X$ and could be smaller than log of the dimension of the representation $R$.
Then the entropy of the interior modes exceeds the area of the almost evaporated black hole
 \be 
\frac{A}{4G_{\text N}} = \pi X^2 <{\log \dim R} < S_\text{interior}\,.
 \ee
 The covariant entropy bound \cite{Bousso:1999xy} states that the matter entropy on a lightsheet is bounded by the change of the transverse area: $ S_\text{lightsheet} \leq \frac{\Delta A}{4G_{\text N}}$.
  Assuming that the matter entropy on the lightsheet bounds the matter entropy in the black hole interior $S_\text{lightsheet} \sim S_\text{interior}$ (we will revisit this point in the end of this section), this leads to a contradiction.

 The above argument can be improved to include the cases
 when the group $G$ is 
 abelian or when it does not have representations with large dimensions.
 One instead considers forming black holes with several possible representations $R \in \mathcal R$, for some large set of unitary irreducible representations $\mathcal R$. As explained before, since Hawking radiation is thermal, we can assume that the black holes maintain their representations $R$ throughout their evaporation process. When all black holes reach the mass $M = X M_{\text pl}$, we thus obtain a number $\sum_{R\in \mathcal R} \dim R$ of remnant states that are indistinguishable. Once again, if $ \pi X^2< \log \sum_{R\in \mathcal R} \dim R$, making the same set of assumptions as above, the existence of such objects is inconsistent with the covariant entropy bound. In other words, the presence of remnants, whose entropy can be arbitrarily large due to the presence of global symmetry, is
 inconsistent with the (naive use of) covariant entropy bound. Consequently, this stronger version of the above argument rules out the existence of any global symmetry with Lie group $G$ or any finite global symmetry $G$ with large enough unitary representations. 
 
As previously hinted, the above arguments require several technical assumptions. The main technical assumption is that the entropy on the lightsheet can be related to the entropy on some space-like Cauchy slice stretching through the interior of the black hole/remnant. This relation is unclear when the lightsheet  (drawn in orange in figure \ref{fig:Penrose Diagram}) intersects the singularity of the black hole (the red line) \cite{Bousso:2014sda}. Thus, it is unclear whether the covariant entropy bound actually applies to the matter entropy on the whole interior slice of the black hole (or remnant).\footnote{There have been numerous other arguments against the existence of remnants \cite{Giddings:1993km,Susskind:1995da}, however they require different technical assumptions which we do not address in this paper. } 
Rather, it should be the ``central dogma'' \cite{Almheiri:2020cfm} described in the introduction that bounds the number of states inside the black hole interior, at least seen from outside. 
Furthermore, the argument that the entropy of the black hole exceeds the dimension of certain representations can be made only at late-times when the black hole has almost fully evaporated. It would therefore be interesting to understand whether recently discussed effects coming from the contribution of replica wormholes (which completely alter late-time observables, such as the entanglement entropy of the Hawking radiation \cite{Almheiri:2019qdq, Penington:2019kki} or correlators of matter fields \cite{Saad:2019pqd}), affect the conclusions of \cite{Banks:2010zn}. We will explicitly address this in section \ref{sec:state-reconst} and \ref{sec:BH-remn}.

  \subsection{The Harlow-Ooguri and Harlow-Shaghoulian argument}
  
  The holographic principle, or the AdS/CFT correspondence, provides well-defined quantum gravity theories from their boundary dual descriptions.
  In such context, Harlow and Ooguri construct a new argument against global symmetries in gravity using the idea of the entanglement wedge reconstruction \cite{Harlow:2018jwu,Harlow:2018tng}.
  Recently, Harlow and Shaghoulian \cite{Harlow:2020bee} extended this argument to more general evaporating black holes based on the recent development of the Page curve that describes the evolution of the entanglement entropy of Hawking radiation \cite{Penington:2019kki, Almheiri:2019qdq, Penington:2019npb, Almheiri:2019psf, Almheiri:2019hni}.
  They imagine a setup in which the spacetime of an evaporating black hole is separated into two regions $S$ and $R$: $S$ contains the black hole and can be understood by its boundary description, while $R$ stands for a ``reservoir'' absorbing the Hawking radiation and is understood as the exterior region where gravity effects are nonessential.
  The Page curve describes the entropy of the Hawking radiation in $R$.
  The entropy of the radiation first grows due to the thermalization between $S$ and $R$ and is bounded by the Bekenstein-Hawking entropy of the black hole system, that means it should decrease after the Page time when it saturates the black hole entropy.
  In the gravity picture, such a transition corresponds to a phase transition of two Quantum Extremal Surfaces (QES) \cite{Engelhardt:2014gca}: one is the empty surface and the other is close to the black hole horizon.
  After the Page time, the nontrivial QES will dominate and enclose a large portion of the black hole interior called the island, which belongs to the entanglement wedge of $R$ but not $S$.
  On the other hand, if one considers a smaller portion of $R$ when the naive thermal entropy of the system does not exceed the black hole entropy, there will be no Page transition and therefore no island.
  Based on this, Harlow and Shaghoulian argue that the unitary transformation generated by the global symmetry group can be split into products of unitary transformations on $S$ and small portions $R_i$ of $R$ ($\cup_i R_i=R$):
  \be
  U(g)=U(g,S)\prod_i  U(g, R_i) U_\text{edge}
  \ee
  where $U_\text{edge}$ is only supported on the edges of $R_i$.
  After the Page time, the island region is not contained in any of the entanglement wedge of $S$ and $R_i$'s. Such a global symmetry cannot act on any simple operators in the island, and therefore the global symmetry cannot exist. 
  
  Given that the island formula has recently been ``derived'' from the contribution of replica wormholes to the gravitational path integral \cite{Penington:2019kki,Almheiri:2019qdq}, it would be informative to understand why the global symmetry cannot exist without relying on the existence of islands. In the next sections we will directly address what effects Euclidean wormholes have on global symmetries present in quantum gravity.

  \section{Global symmetry violation from replica wormhole}
  \label{sec:replica-wormhole-and-global-symm}
  
  \subsection{General argument}
  \label{sec:general-arg}
  
In this section, we will provide a new argument about the nonexistence of exact global symmetry in quantum gravity.
We will argue that even if the low-energy effective action of the theory preserves some global symmetry, the global symmetry charge conservation is violated in quantum gravity due to the existence of replica wormholes.
More precisely, the replica wormhole will predict a nonzero transition probability between states with different symmetry global charges.

In order to understand the contribution of such replica wormholes, we should first list the necessary assumptions for computing observables in a gravitational theory.  Throughout this paper we will be interested in preparing states in some gravitational theory, with metric $g_{\mu \nu}$, coupled to a
matter field $\Phi$, such that the  Lagrangian $\mathcal{L}(\Phi, g_{\mu \nu})$ of the theory  is invariant under some global transformation $G$.
We can prepare such a state using the Euclidean path integral, with Dirichlet boundary conditions for the metric ({\it i.e.}~the Hartle-Hawking state $|HH\rangle$),\footnote{We would like to clarify that this is the Hartle-Hawking state describing the thermal state of a black hole \cite{PhysRevD.13.2188}, not the Hartle-Hawking no-boundary state \cite{PhysRevD.28.2960}.} which means that  boundary operator insertions can be specified in a diffeomorphism invariant way. Thus, for some operator $\cO(\tau)$ formed from the matter field $\Phi$, we can define some state $|\psi\>$ as,
  \be
  \label{eq:state-prep}
  |\psi\> = \cO(\tau) |HH\> =  \begin{tikzpicture}[baseline={([yshift=-.5ex]current bounding box.center)}, scale=0.6]
 \pgftext{\includegraphics[scale=0.25]{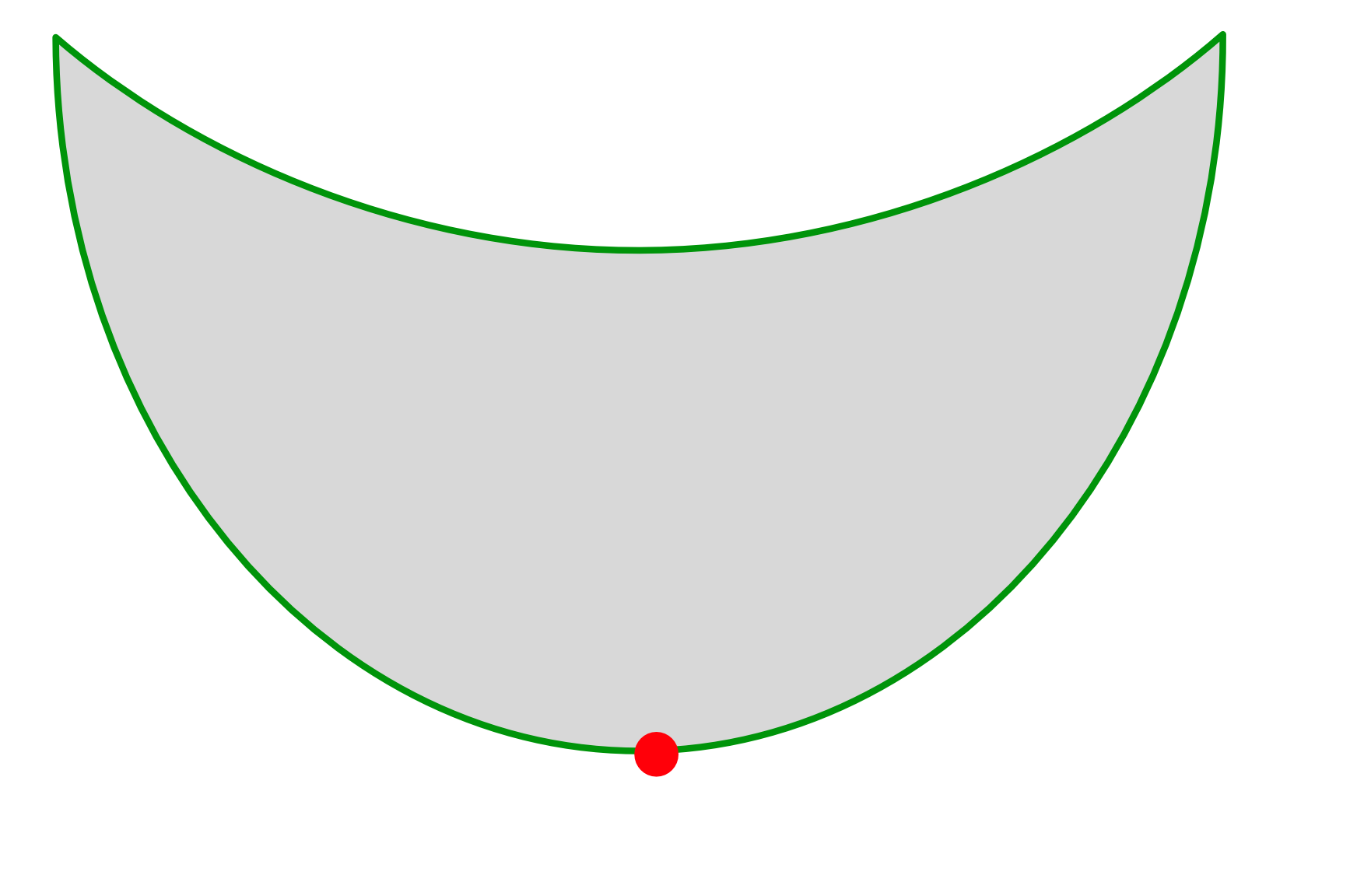}} at (0,0);
   \draw (0,-2.3) node  {$\mathcal O(\tau)$}; 
  \end{tikzpicture}
  \,.
  \ee
  
 The excitations produced by the matter field $\Phi$ can be classified according to their representations of $G$.
For example, consider two operators ${\cal O}_1,{\cal O}_2$ transforming under representation $R_1$ and $R_2$ of a global symmetry $G$, which, when acting on the Hartle-Hawking state, create two states $|R_1\>$ and $|R_2\>$  with representation $R_1$ and $R_2$. A nonzero transition amplitude between these two states $\langle R_2|R_1\>=\langle{\cal O}_2^\dag{\cal O}_1\rangle$ would imply that the global symmetry $G$ is broken, given that $\olsi{R}_1\otimes R_2$ does not contain the singlet. The physical observable constructed from the transition amplitude between the two states is the scattering probability $  |\< R_2|R_1\>|^2$. For simplicity, we will for now consider the case in which there is no Lorentzian evolution between the in- and out-states and when the global symmetry $G$ is never spontaneously broken.\footnote{Even if $G$ is spontaneously broken we can choose appropriate boundary condition to make the charged field to have vanishing expectation value.
We will demonstrate this through an example in section \ref{sec:JT+gauge-theory}. } Without Lorentzian evolution, the scattering probability yields the squared norm of the inner-product between the two states which, due to the charge conservation, would simply vanish in quantum field theory. This will not be the case when coupling $\Phi$ to gravity.

In terms of the gravitational path integral, the scattering probability  can be written as
  \be 
  \label{eq:out-in-1}
  |\< R_2|R_1\>|^2 = \tr \rho_{\text{in-in}} \rho_{\text{out-out}} = \begin{tikzpicture}[baseline={([yshift=-.5ex]current bounding box.center)}, scale=0.6]
 \pgftext{\includegraphics[scale=0.35]{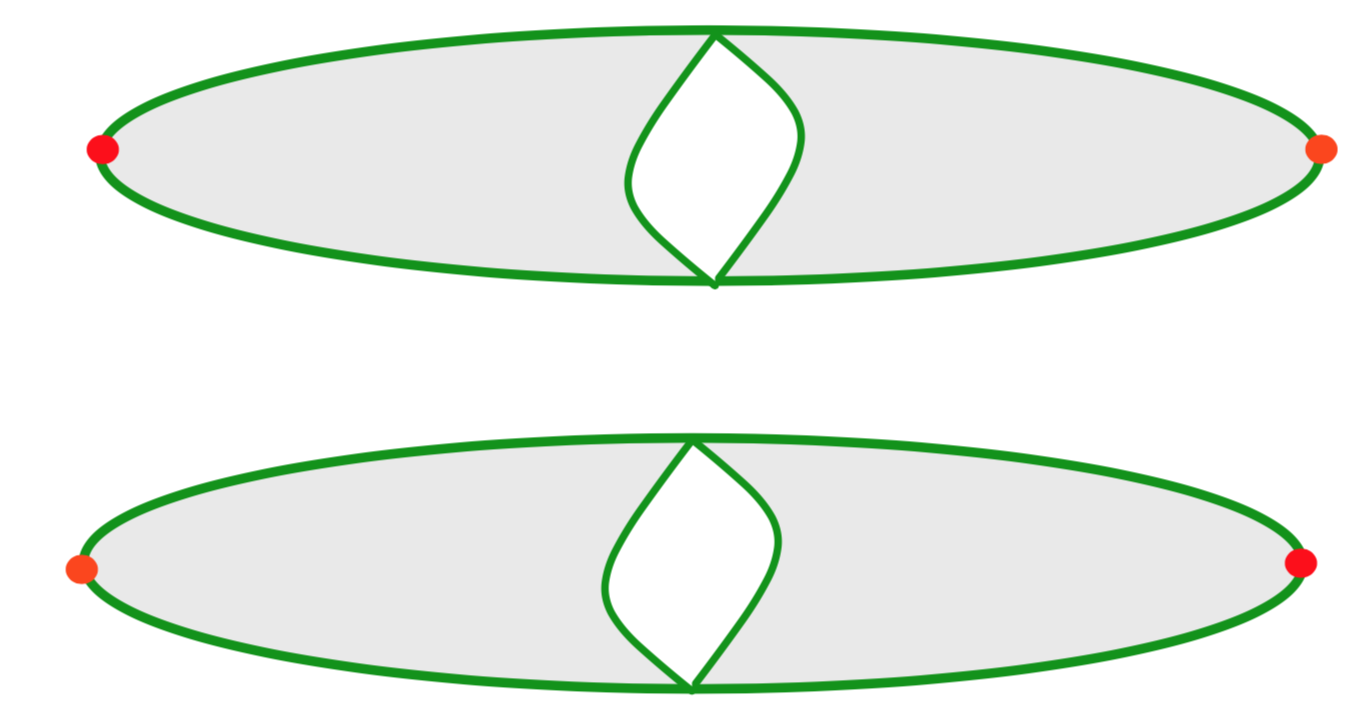}} at (0,0);
   \draw (-4.0,-1.7) node  {$\mathcal O_1$}; 
     \draw (4.5,-1.7) node  {$\mathcal O_2$}; 
        \draw (-4.0,1.85) node  {$\mathcal O_1^\dagger$}; 
     \draw (4.5,1.85) node  {$\mathcal O_2^\dagger$}; 
    \draw (0,0) node  {\huge{$\dots$}}; 
  \end{tikzpicture}
  \ee
  where when rewriting this probability in terms of two density matrices, $ \rho_{\text{in-in}}=|R_1\>\<R_1|$ and $\rho_{\text{out-out}}=|R_2\>\< R_2| $, we emphasize that just like when computing the Renyi entropies in a gravitational theory  \cite{Penington:2019kki, Almheiri:2019qdq}, we need to consider several (for the scattering probability, only two) replicas  of the gravitationally prepared density matrices. Furthermore, this rewriting  emphasizes  the relation between this probability and the $\$$-matrix considered in section \ref{sec:hawking-arg}. 
  
  Next, we assume that between the two replicas in \eqref{eq:out-in-1}, there is no restriction on the Euclidean gravitational path integral which would disallow connected geometries.\footnote{While we do not yet know of a reason whether to include (or exclude) connected replica geometries in a UV completion of gravity, from  the path integral perspective, there is no way to impose that we only sum over connected geometries when using a local measure for the metric $g_{\mu \nu}$.} Thus, the inner-product between the states $|R_1\>$ and $|R_2\>$ is given by 
    \be 
    \label{eq:in-out-2}
  |\< R_2|R_1\>|^2 = \tr \rho_{\text{in-in}} \rho_{\text{out-out}} &= \underbrace{
  \begin{tikzpicture}[baseline={([yshift=-.5ex]current bounding box.center)}, scale=0.45]
 \pgftext{\includegraphics[scale=0.2]{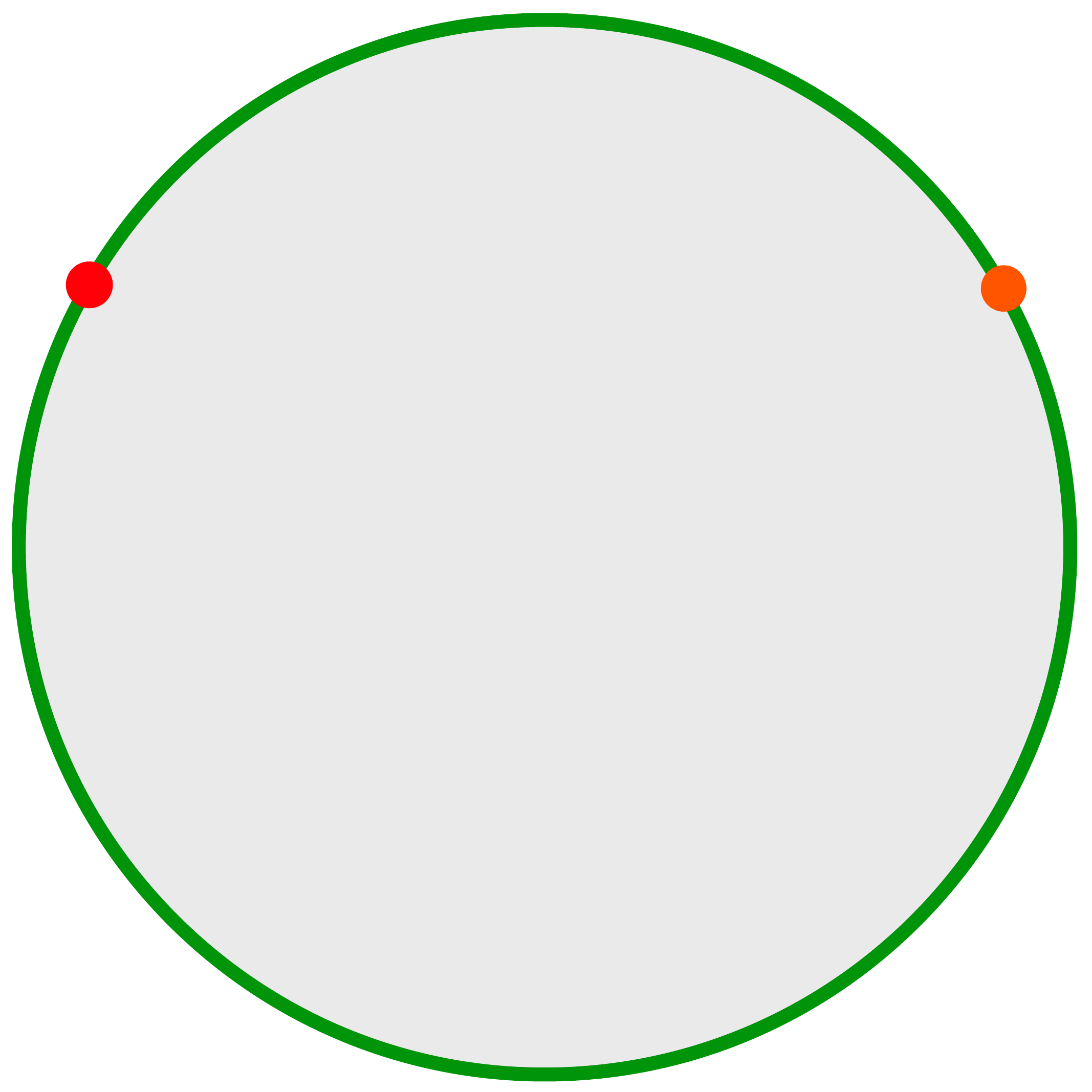}} at (0,0);
 \draw (-2.99,+1.05) node  {$\mathcal O_{R_1}$};
  \draw (2.95,+1.05) node  {$\mathcal O_{R_2}^\dagger$};
  \end{tikzpicture}
  \hspace{0.3cm}
 \begin{tikzpicture}[baseline={([yshift=-.5ex]current bounding box.center)}, scale=0.45]
 \pgftext{\includegraphics[scale=0.2]{disk2.pdf}} at (0,0);
 \draw (-3.2,+1.05) node  {$\mathcal O_{R_1}^\dagger$};
  \draw (2.95,+1.05) node  {$\mathcal O_{R_2}$};
  \end{tikzpicture}
  }_{\text{Vanishing contribution}} \nn 
  \,\,\, \\ 
   &\hspace{-2.8cm}+  \underbrace{\begin{tikzpicture}[baseline={([yshift=-.5ex]current bounding box.center)}, scale=0.45]
 \pgftext{\includegraphics[scale=0.35]{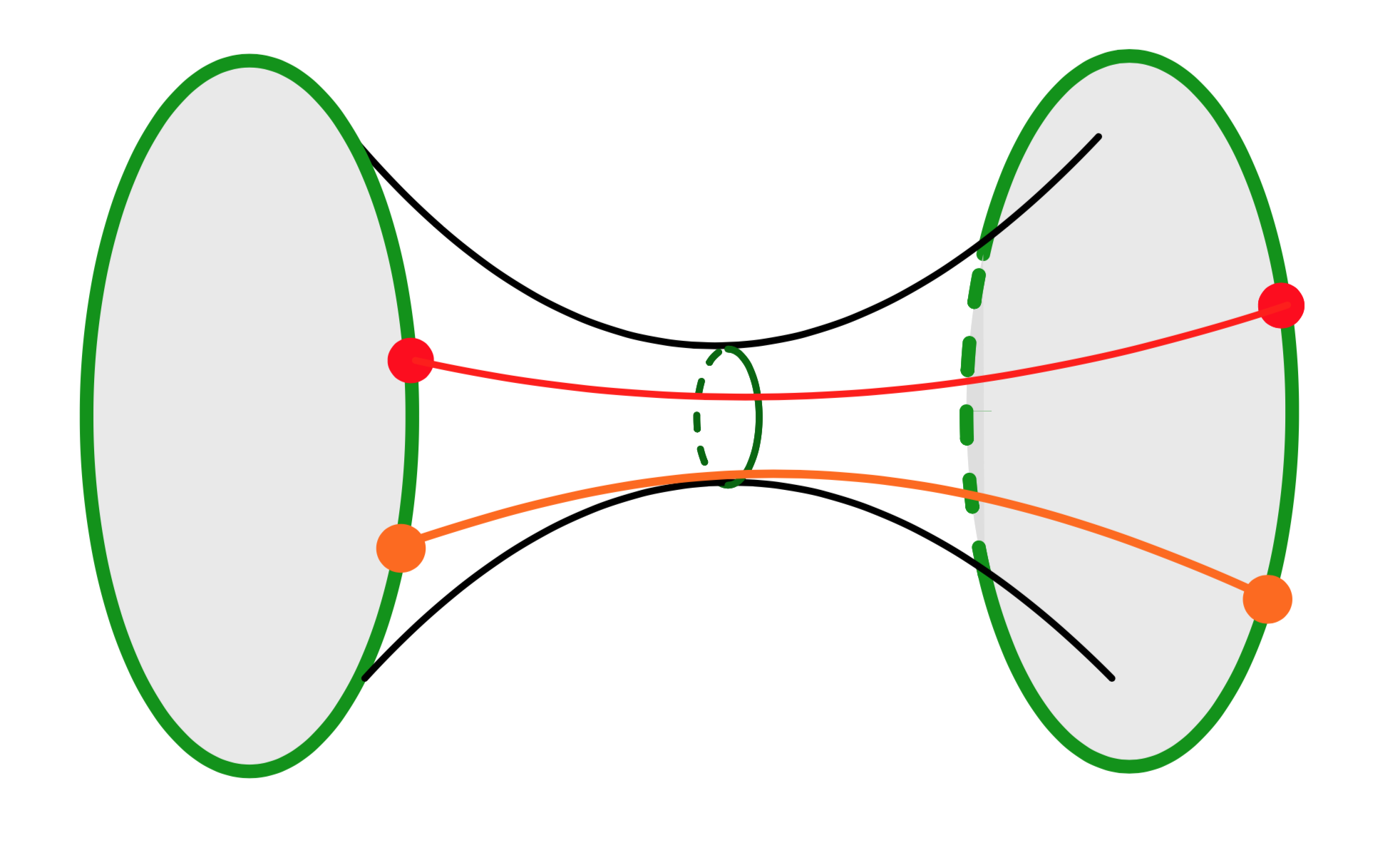}} at (0,0);
 \draw (-3.5,+1.05) node  {$\mathcal O_{R_1}$};
 \draw (-3.5,-1.05) node  {$\mathcal O_{R_2}^\dagger$};
 \draw (6.0,+1.05) node  {$\mathcal O_{R_1}^\dagger$};
 \draw (6.0,-1.05) node  {$\mathcal O_{R_2}$};
  \end{tikzpicture}}_{\text{Leading contribution}}
   +  \underbrace{\dots}_{\text{Higher genus sub-leading contributions}}\,.
  \ee
  The leading disconnected geometries are given in the first line and their contribution is an integral over the correlators $\<\cO_1^\dagger \cO_2\>$ and $\<\cO_1\cO_2^\dagger\>$ on two disconnected fluctuating geometries. However, since we have assumed that $\bar R_1 \times R_2$ contains no singlets, the correlators $\<\cO_1^\dagger \cO_2\>$ and $\<\cO_1\cO_2^\dagger\>$ vanish on all backgrounds due to charge conservation, when assuming that the global symmetry $G$ is not spontaneously broken. The second line in \eqref{eq:in-out-2} yields the contribution of the replica wormhole ({\it i.e.}~the connected geometry) and its contribution is given by the correlator $\<\cO_1^\dagger \cO_2 \cO_1 \cO_2^\dagger \>$ evaluated on a sum over fluctuating connected geometries. This correlator is generically non-vanishing on any geometry since singlets are always present in the tensor products $\bar R_1 \times R_1$, and $\bar R_2 \times R_2$ (following from the definition of the complex conjugate irreducible representation). The connected geometry yields a non-zero contribution and is especially trustworthy, even in a theory whose UV completion is unknown, when the replica wormhole geometry (in the presence of operator insertions) is a saddle point of the gravitational path integral. We will explicitly compute the value of this correlator in the next subsections in two simple toy-models: in section \ref{sec:JT+matter}, in JT gravity coupled to a massive scalar field theory with a $U(1)$ global symmetry, where we will show that the wormhole is indeed a saddle for the black hole geometry, and, in section \ref{sec:JT+gauge-theory}, in a simpler two-dimensional topological theory of gravity, coupled to a $\mathbb Z_k$ gauge theory (which can be expressed as a $U(1)\times U(1)$ one-form gauge field and periodic scalar \cite{Maldacena:2001ss,Banks:2010zn,Kapustin:2014gua}). 
   The only other contributions to the inner-product are given by sub-leading, typically higher topology, geometries which are non-perturbatively suppressed either because they capture the contribution of sub-leading saddles or, as is the case for JT gravity \cite{Saad:2019lba}, because they are exponentially suppressed by $e^{-2(\text{BH entropy}) g}$ where $g$ is the genus of the connected manifold. \vspace{-0.2cm}
  
  To obtain the scattering probability, we need to analytically continue the geometries above, to have a period of Lorentzian evolution between the in- and out-states. Such an analytic continuation is schematically given by, \\\vspace{-0.2cm}
\be 
\label{eq:in-out-3}
 |\<  R_2|e^{iHt}|R_1\>|^2 =  \tr \rho_{\text{in-in}} e^{i H t}\rho_{\text{out-out}} e^{-i H t} &=  \underbrace{\begin{tikzpicture}[baseline={([yshift=-.5ex]current bounding box.center)}, scale=0.36]
 \pgftext{\includegraphics[scale=0.25]{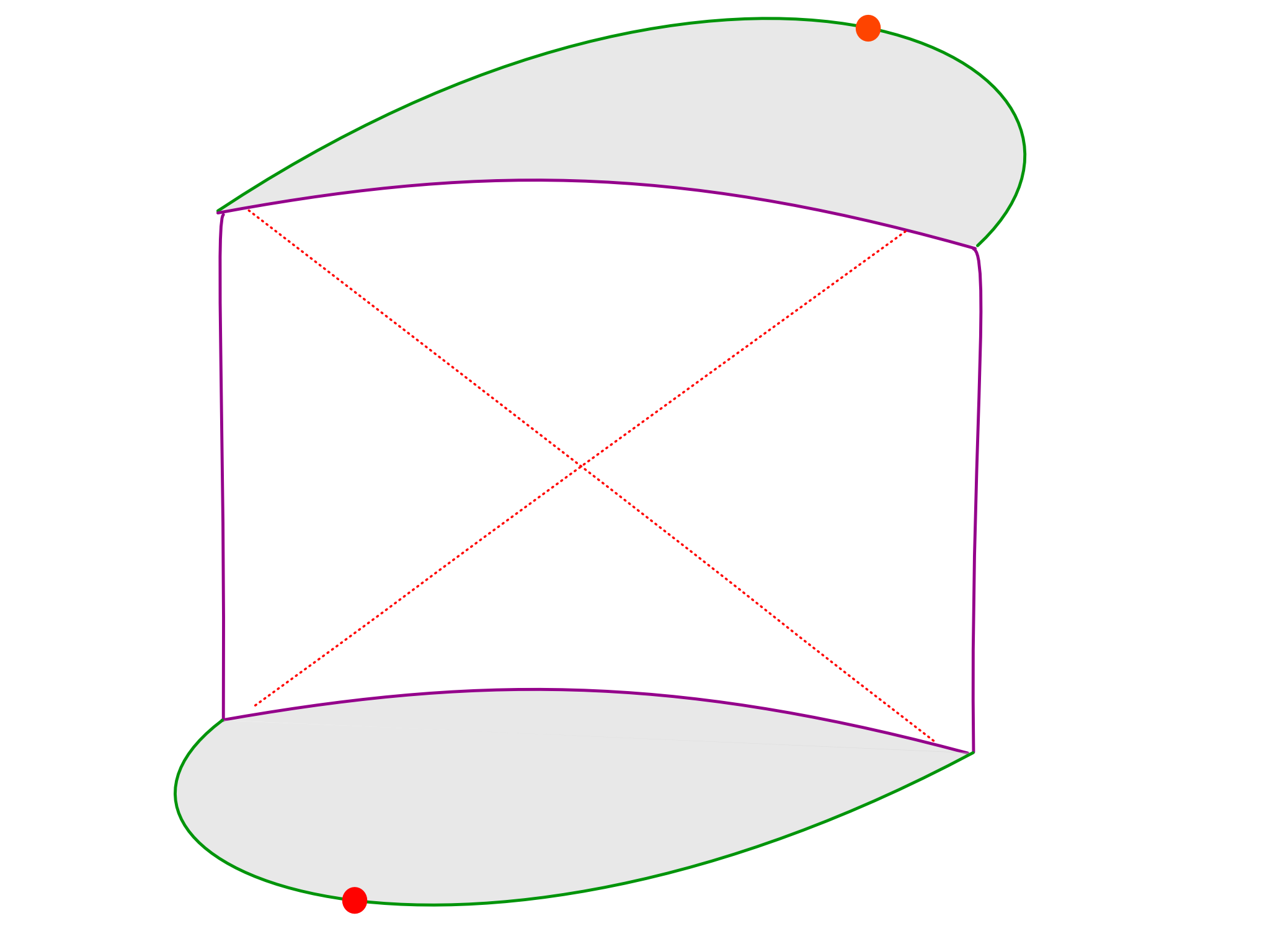}} at (0,0);
\draw (-2.55,-3.2) node  {$\mathcal O_{R_1}$};
  \draw (2.8,+3.2)  node  {$\mathcal O_{R_2}^\dagger$};
  \end{tikzpicture}
 \begin{tikzpicture}[baseline={([yshift=-.5ex]current bounding box.center)}, scale=0.36]
 \pgftext{\includegraphics[scale=0.25]{state_Lorentzian.png}} at (0,0);
\draw (-2.55,-3.2) node  {$\mathcal O_{R_1}^\dagger$};
  \draw (2.8,+3.2)  node  {$\mathcal O_{R_2}$};
  \end{tikzpicture}}_{\text{Vanishing contribution}} \nn 
  \,\,\, \\ 
   & \hspace{-5.5cm} + \underbrace{\begin{tikzpicture}[baseline={([yshift=-.5ex]current bounding box.center)}, scale=0.36]
 \pgftext{\includegraphics[scale=0.45]{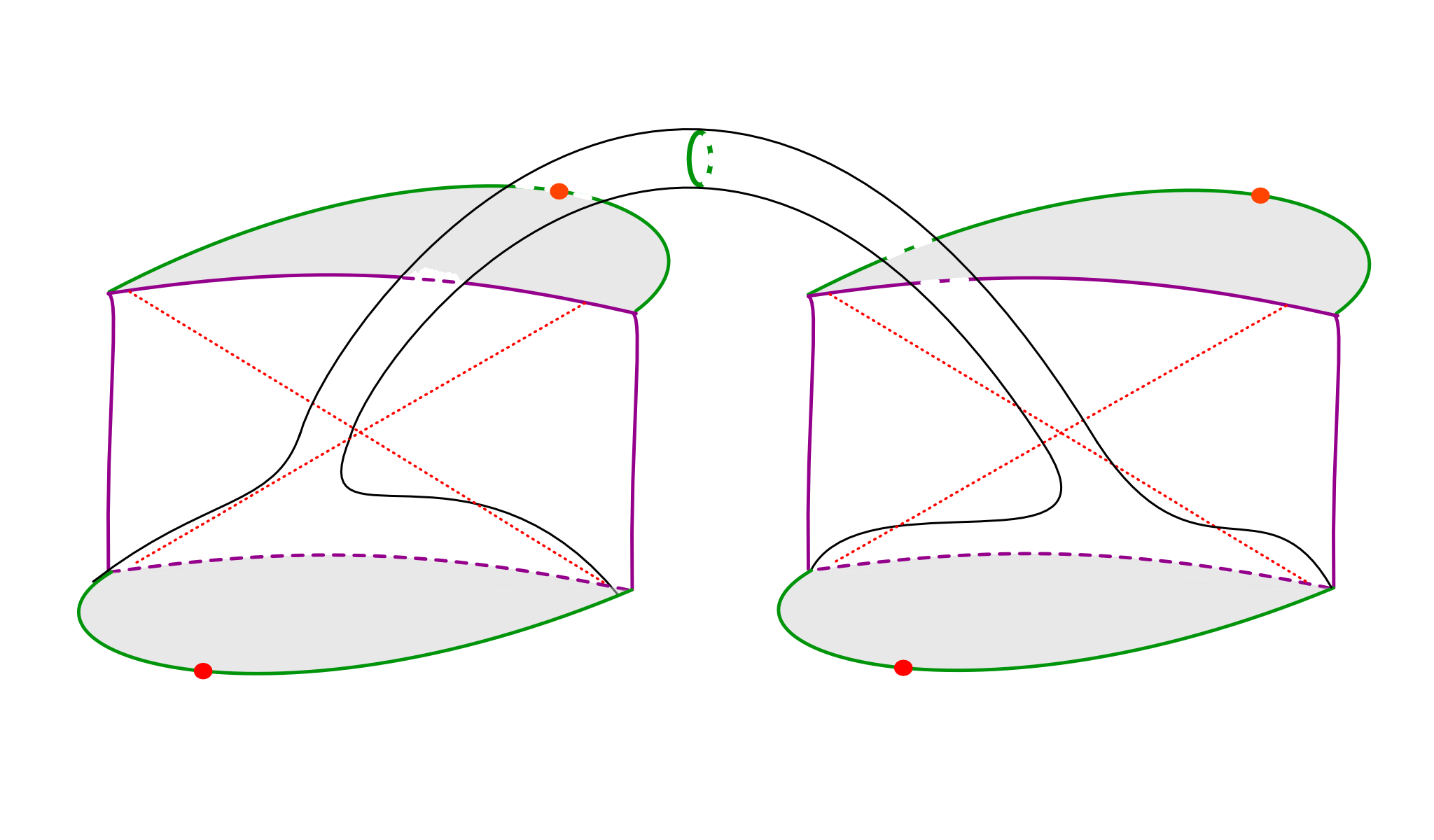}} at (0,0);
 \draw (-2.8,+3.25) node  {$\mathcal O_{R_2}^\dagger$};
 \draw (-5.2,-3.5) node  {$\mathcal O_{R_1}$};
 \draw (5.2,+3.25) node  {$\mathcal O_{R_2}$};
 \draw (3.4,-3.5) node  {$\mathcal O_{R_1}^\dagger$};
  \end{tikzpicture}}_{\text{Leading contribution}} +  \underbrace{\dots}_{\text{Higher genus sub-leading contributions}}\,.
  \ee
  The first line, again shows the contribution of the vanishing disconnected geometries on the black hole background, whose horizon  is shown by the red dotted lines and where periods of Lorentzian evolution are symbolized by green curves and periods of Lorentzian evolution are symbolized by purple boundary curves. The Lorentzian replica wormhole contribution is captured on the second line of \eqref{eq:in-out-3}, where the order in which the Euclidean patches are connected is unimportant.  Notice that again, since the scattering probability is given by a simple analytic continuation of the inner-product in \eqref{eq:in-out-2},  although the wormhole is an instanton contribution to the gravitational path integral and is $e^{-{1\over G_{\text N}}}$ suppressed, it dominates over the original geometry which gives a vanishing answer.
Therefore we conclude that the transition probability, just like the inner-product norm, does not vanish and the charge conservation of the global symmetry $G$ is violated in quantum gravity, even when there is no explicit breaking of the symmetry in the Lagrangian.
We can in principle consider more general density matrices $\rho_{\text{in-in}}$ and $\rho_{\text{out-out}}$ and the comments regarding the contributions of replica wormholes will still follow as long as the path integral on some generic wormhole geometry is non-vanishing. 

Finally, if the bulk theory has a boundary dual, where we have to consider an ensemble average of theories due to the wormholes, the nonzero transition probability between charged states in the bulk means that the corresponding inner-product between states on the boundary is some nonzero random number.
After taking the ensemble average of a single copy of such a system, this inner-product vanishes, while if taking two copies of the system, needed to compute the absolute value of the inner-product, we find a non-vanishing answer. 
As previously described, this implies that the system develops a global symmetry only after taking the ensemble average and we will discuss examples in section \ref{sec:how-global-symm-can-arise}. 
This provides us a new aspect of the issue of factorization: if we want to restore the factorization in the bulk with a modifications of the Lagrangian of the theory, 
this inevitably requires an explicit breaking of all the global symmetries to be consistent with the wormhole calculation. We will discuss this point further in section \ref{sec:how-global-symm-can-arise}.

\subsection{A few examples}
In this section, we will provide a few examples that support our argument. The main examples are based on two dimensional gravitational theories where the wormhole configurations are best understood. However, we except the argument can extend to higher dimensions, where wormhole geometries can also be constructed \cite{Maldacena:2004rf}.

   \subsubsection{JT gravity coupled to matter}
  \label{sec:JT+matter}

    \begin{figure}[t!]
  \begin{center}
 \begin{tikzpicture}[scale=1.0]
 \pgftext{\includegraphics[scale=0.2]{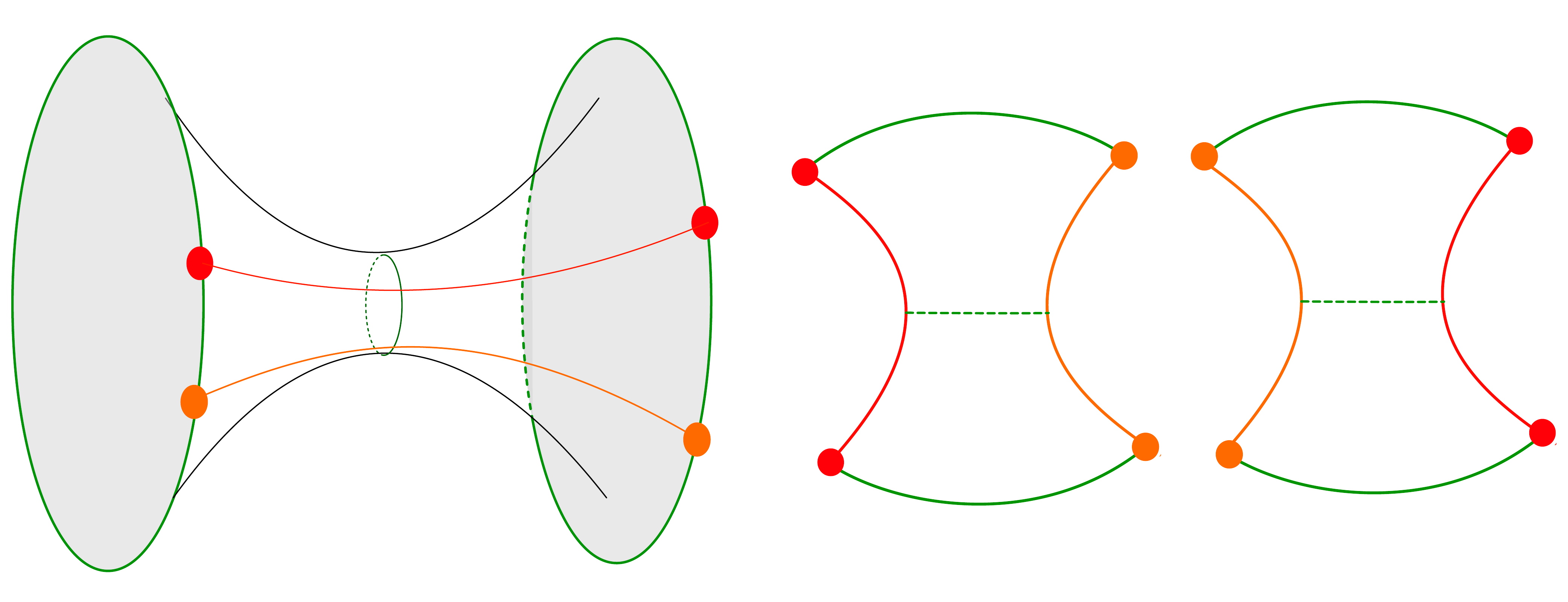}} at (0,0);
   \draw (-0.0, 0) node  {$=$};
    \draw (3.5, 0) node  {$+$};
  \end{tikzpicture}
 \caption{\label{fig:decomp-wormhole}A decomposition of the Euclidean wormhole seen in the second line of \eqref{eq:in-out-2} into patches on the Poincar\'e disk.}
  \end{center}
  \end{figure}
  
  Our first example is JT gravity \cite{Jackiw:1984je,Teitelboim:1983ux} coupled with bulk matter field which has a global symmetry $G$. 
Such a theory is not UV complete due to the contribution of higher topology geometries to the path integral \cite{Saad:2019lba}. To estimate the effect of such geometries, we use the semi-classical gravitational saddles.
For concreteness and simplicity, we can take the matter field to be a complex scalar $\varphi$ which carries some charge $q$ under a $U(1)$ subgroup of $G$.
Consider two charge states $|q\rangle$ and $|\bar q\rangle$ prepared by insertion of $\varphi$ and $\varphi^*$ on a thermofield double state with inverse temperature $\beta$, prepared as described in section \ref{sec:general-arg}:
\be
\begin{split}
|q\rangle_{\beta}=\sum_{n}e^{-{\beta\over 4}H_L-{\beta\over 4}H_R}\varphi_L|n\rangle_L|n\rangle_R\,,\qquad 
|\bar q\rangle_{\beta}=\sum_{n}e^{-{\beta\over 4}H_L-{\beta\over 4}H_R}\varphi^*_L|n\rangle_L|n\rangle_R\,,
\end{split}
\ee
prepared as in \eqref{eq:state-prep}.

The inner-product between these two states is equal to the two-point function 
\begin{eqnarray}
\langle \bar q|q\rangle_{\beta} =\<\varphi(\beta/2) \varphi(0)\>_{\beta}=\Tr \left(e^{-{\beta\over 2} H}\varphi e^{-{\beta\over 2} H}\varphi\right)
\end{eqnarray}
on a single boundary.
In JT gravity, such a correlation function is given by a summation over all hyperbolic geometries ending on the single boundary, including the disk topology and adding handles on it.
As we discussed in the section above, the two-point function $\<\varphi(\beta/2) \varphi(0)\>_{\beta}$ on such geometries is identically zero, due to the existence of the U(1) symmetry.

let us now consider the absolute value squared of the inner-product, $|\<\bar q|q\>|^2$ as in \eqref{eq:in-out-2}. 
In JT gravity, such a two boundary correlator is given by summation over all smooth hyperbolic geometries ending on the two boundaries, including factorized geometries and non-factorized geometries.
The correlator on the factorized geometry is zero, since there will be two independent U(1) global symmetry acting on the two boundaries and for each boundary the insertion of the operators ($\varphi \varphi$ and $\varphi^*\varphi^*$) is not $U(1)$ invariant.
On the other hand, the correlator on the non-factorized geometry can generically be nonzero, since now there will be only one unique global $U(1)$ symmetry acting on these two boundaries, and the whole quantity is invariant under such global transformations.

Our next step is then to construct such a semi-classical wormhole geometry. The leading wormhole geometry that connects between these two asymptotic boundaries is the double-trumpet geometry (the second line in  \eqref{eq:in-out-2}).
On such geometries, there exist non-vanishing bulk propagators connecting $\varphi$ to $\varphi^*$ on the two sides.\footnote{
Such a quantity has recently been considered by Stanford \cite{Stanford:2020wkf} and is further reviewed in appendix \ref{app:one and two-point function} using the exact quantization method of JT gravity.
}
In JT gravity, such a configuration can be evaluated directly by cutting the double trumpet along the geodescis connecting $\phi$ and $\phi^*$, along which the particle propagate semi-classically when their masses $m$ is sufficiently large (see figure \ref{fig:decomp-wormhole}).
Denoting the geodesic distance along the two geodesics as $\ell_{1,2}$, the propagator can be approximated as $e^{-m\ell_{1,2}}$.
The gravitational path integral on the rectangular region with two asymptotic boundaries with length $\beta_{L,R}$ and two geodesic lengths $\ell_{1,2}$ is then given by \cite{Mertens:2017mtv, Kitaev:2018wpr, Yang:2018gdb, Blommaert:2018oro, Iliesiu:2019xuh,Saad:2019pqd,Stanford:2020wkf}:
\be
\langle \ell_1|e^{-\beta_L H_L-\beta_R H_R}|\ell_2\rangle=\int \mathrm{d}E \rho(E) e^{-(\beta_L+\beta_R)E}\langle \ell_1|E\rangle\langle E|\ell_2\rangle\,,
\ee
where $\rho(E)={1\over 2\pi^2}\sinh(2\pi\sqrt{2E})$ is the density of states and $\langle \ell|E\rangle=4K_{2i\sqrt{2E}}(4 e^{-\ell/2})$ is the Wheeler-DeWitt wavefunction in the energy basis \cite{Kitaev:2018wpr, Yang:2018gdb}.
For the moment, we will focus on the partition function with matter field ignored, which we will revisit later.
Then the whole path integral can be obtained by gluing these two rectangular regions with the propagator:
\be
\begin{split}
\begin{tikzpicture}[baseline={([yshift=-.5ex]current bounding box.center)}, scale=0.35]
 \pgftext{\includegraphics[scale=0.35]{cylinder2.png}} at (0,0);
 \draw (-3.2,+1.05) node  {$\phi$};
 \draw (-3.2,-1.05) node  {$\phi$};
 \draw (4.2,+1.05) node  {$\phi^*$};
 \draw (4.2,-1.05) node  {$\phi^*$};
  \end{tikzpicture} &=\int \mathrm{d}\ell_1 \mathrm{d}\ell_2 \langle \ell_1|e^{-\beta_L H_L-\beta_R H_R}|\ell_2\rangle \langle \ell_1|e^{-\beta_L' H_L-\beta_R' H_R}|\ell_2\rangle e^{-m\ell_1-m\ell_2}\\
&\hspace{-2.5cm}=\int \mathrm{d}E_1 \mathrm{d}E_2 \rho(E_1)\rho(E_2) e^{-(\beta_{L}+\beta_{R})E_1-(\beta_{L}'+\beta_{R}')E_2}\langle E_1|\mathcal{O}^{\dagger}\mathcal{O}|E_2\rangle\langle E_2|\mathcal{O}^{\dagger}\mathcal{O}|E_1\rangle\,,
\end{split}
\ee
where $\langle E|\mathcal{O}^{\dagger}\mathcal{O}|E'\rangle$ is the two-point function in energy basis on a disk \cite{Mertens:2017mtv, Kitaev:2018wpr, Yang:2018gdb, Blommaert:2018oro, Iliesiu:2019xuh}:
\begin{align}
\langle E|\mathcal{O}^{\dagger}\mathcal{O}|E'\rangle & = {1\over 2^{2m+1}\Gamma(2m)} \Gamma(m+ i(\sqrt{2E}+\sqrt{2E'})\Gamma(m- i(\sqrt{2E}+\sqrt{2E'}) \nn \\ & \qquad \qquad \,\,\, \,\,\,\times \Gamma(m+ i(\sqrt{2E}-\sqrt{2E'})\Gamma(m- i(\sqrt{2E}-\sqrt{2E'})\,\nn \\ 
&\equiv {\Gamma(m\pm i(\sqrt{2E}\pm \sqrt{2E'}))\over 2^{2m+1}\Gamma(2m)} ~,
\end{align}
where the $\pm$ sign in the last line above means we take the product of all the four gamma functions coming from different choices of the $\pm$ signs.

In the symmetric configuration where $\beta_L+\beta_R=\beta_L'+\beta_R'=\beta$,  $E_{1,2}$ has the same saddle point $E$.
The action of $E$ contains two pieces: the gravitational action contributes the usual thermal action, $S(E)-\beta E$, and the propagator contributes an action of order $m\log E-S(E)$ coming from the asymptotic expansion of the gamma functions.

Together, this leads to a semi-classical saddle of the energy:
\be
E_\text{saddle}={m\over \beta}.
\ee
Now, let us examine our assumption of ignoring the matter partition function.
In order to justify that, we need the size of the wormhole, $b$ ({\it i.e.}~the geodesic length across the wormhole), to be large in order to ignore possible bulk matter excitations and their back-reaction to the geometry \cite{Saad:2019lba}. If that were not the case, the matter partition function would have a divergence for small values of $b$.
In appendix \ref{app:one and two-point function}, this size has been estimated directly using the cross ratio of the four corners of the rectangular region.
As a result, when we do a Lorentzian time evolution $\beta_{L,R}\rightarrow \beta_{L,R}\pm i T$ and $\beta'_{L,R}\rightarrow \beta'_{L,R}\mp i T$, the size of the wormhole grows linearly with time:
\be
b \sim 2\sqrt{2E_\text{saddle}} T\,.
\ee
Therefore, as long as $T$ is large enough, we can ignore the contribution of the matter partition function. The wormhole contribution is then given by:
\be
|\< \bar q|q\>_{\beta}|^2\sim {1\over |\< q|q\>_{\beta}|^2}({8m\over  \beta})^{2m} e^{-2m} {\Gamma(m)^4\over 2^{4m}\Gamma(2m)^2} \sim ({\beta m\over 2\pi^2})^{2m} e^{-2S_0-\frac{4\pi^2}{\beta}-2m}\,,
\ee
where we assume $m$ is large. $\< q|q\>$ is a normalization factor, and is given by the disk two-point function:
\be
\< q|q\>_{\beta}\sim {\Gamma(m)^2\over 2^{2m}\Gamma(2m)}({4\pi\over \beta})^{2m} e^{S_0+2\pi^2\over \beta} .
\ee

Finally, let us remark on the situation when one of the operators whose scattering probability we want to compute is the identity operator. Once again, the leading wormhole is the double-trumpet geometry and the same cutting-and-gluing rule explained above can be used to obtain the expectation value if we ignore the contribution of the matter partition function \cite{Saad:2019pqd}.
The result is similar to the two-point function case:
\be
|\<\varphi\>_{\beta}|^2=\int \mathrm{d}E \rho(E) e^{-(\beta_L+\beta_R) E} \langle E|\mathcal{O}^{\dagger}\mathcal{O}|E\rangle.
\ee
The saddle point discussion is almost identical with the two-point function case, so we will not repeat that analysis here.
Again, to justify the assumption of neglecting the matter partition function, one needs to do the same analytic continuation of $\beta_{L,R} \rightarrow \beta_{L,R}\pm i T$.  
The only difference between this case and the two-point function case is that the total boundary time evolution in this case is mostly Lorentzian, just as the situation of the spectrum form factor.

  \subsubsection{Gravity coupled to a pure gauge theory}
  \label{sec:JT+gauge-theory}
  
  Our second example is 2D gravity coupled to a BF theory \cite{Witten:1991we,Witten:1992xu}, where the 2D gravity could either be pure topological gravity, JT gravity \cite{Saad:2019lba, Iliesiu:2019lfc, Kapec:2019ecr} or its extensions \cite{Maxfield:2020ale, Witten:2020wvy}. The point of considering such a theory is two-fold:
  \begin{itemize}
      \item The first is that the theory is UV complete so we need not worry about the matter partition function divergence coming from wormholes with small size $b$. 
      
      \item  The second is that such a theory has a spontaneous breaking of its zero-form symmetry (whose origin we review below) and we will be able to show that the general analysis in section
     \ref{sec:general-arg} is applicable even in such a case.   
  \end{itemize}
  The action of the theory is \cite{ Iliesiu:2019lfc, Kapec:2019ecr}:
  \begin{equation} 
     S_\text{grav}[g]+\frac{ik}{2\pi}\phi \,\mathrm{d}a+S_\text{boundary}(g,\Phi,a)~,
  \end{equation}
  where $a$ is a $U(1)$ gauge field, and $\phi$ is a scalar with periodicity $2\pi$, which enforces the gauge field $a$ to be flat by the equation of motion.
  $k$ is quantized to be an integer in order for the action to be well-defined for a $2\pi$-periodic scalar $\phi$: under the  ``gauge transformation'' $\phi\rightarrow\phi+2\pi$, the action changes by $ik\int da$. For it to be a multiple of $2\pi i$, $k$ needs to an integer by the Dirac quantization condition $\oint da\in 2\pi\mathbb{Z}$.
  
Let us focus on the BF theory, which is a topological field theory that does not couple to gravity. It describes a $\mathbb{Z}_k$ gauge theory \cite{Maldacena:2001ss,Banks:2010zn,Kapustin:2014gua}.
The theory has the following operators
\begin{equation}
U=e^{i\phi},\quad V=e^{i\oint a},\quad U^k=V^k=1,\quad
UVU^{-1}=e^{2\pi i/k}V~,
\end{equation}
  where the last relation means these operators have non-trivial braiding given by the $k$th root of unity, as can be seen from the Aharonov-Bohm phase.
Thus if we ignore gravity, the global symmetries of the theory are:
\begin{itemize}
    \item \textbf{$\mathbb Z_k$ zero-form symmetry:} the symmetry is generated by $V$, and $U$ carries the unit charge. The symmetry acts as $\phi\rightarrow \phi+\lambda_0$ where $\lambda_0$ is a multiple of $2\pi/k$. This leaves the action invariant using the Dirac quantization condition $\oint \mathrm{d}a\in 2\pi\mathbb{Z}$.\footnote{
In contrast, the gauge transform for the gauge group is $a\rightarrow a+d\lambda'$ that leaves $\phi$ invariant.
If we interpret the $\mathbb{Z}_k$ BF theory as the low-energy theory for a $U(1)$ gauge theory Higgsed to $\mathbb{Z}_k$ by a charge-$k$ scalar, the 0-form global symmetry transforms the $\mathbb{Z}_k$ vortices, where the  $\mathbb{Z}_k$ connection has holonomy $e^{2\pi i/k}$ around the minimal vortex.}
    
    \item \textbf{$\mathbb Z_k$ one-form symmetry:} the one-form symmetry \cite{Gaiotto:2014kfa} is generated by $U$, and $V$ carries the unit charge. 
    The one-form symmetry acts as $a\rightarrow a+\lambda_1$ for one-form $\lambda_1$ with $\mathbb{Z}_k$ holonomy. 
\end{itemize}
The above symmetries should not be confused with the $\mathbb Z_k$ gauge symmetry. 

Within topological field theory in the absence of dynamical gravity, these symmetries are spontaneously broken: for instance, on a spatial circle there are $k$ vacua labelled by  $e^{i\phi}$ which is a $k$th root of unity. On the other hand, when the operator $U$ or $V$ is inserted in a homologically non-trivial cycle, their expectation values vanishes \cite{Gaiotto:2014kfa}.
 
Now, let us consider the theory on a disk $D$ with Dirichlet boundary conditions for the gauge field $a$. The disk partition function for such a theory is given by
\be 
Z_{\text BF}[a] = \sum_{q=1}^k \chi_q(e^{i\oint_{\partial D} a}) \,,
\ee
where each term in the sum corresponds to one of the $k$ vacua and $\chi_q(e^{i\oint_{\partial D} a}) = e^{i q\oint_{\partial D} a}$ is the $\mathbb Z_k$ character with $q \in 1, \dots, k$.

Setting $a =0$, we want to prove that the one-point function of $U$ inserted at the boundary of the disk vanishes, 
\begin{equation}
    \langle U\rangle =0~.
\end{equation}
To see this, we can modify the action by the insertion of $U=e^{i\phi}$ at some point $p$:
\begin{equation}
    \phi(p)+\frac{k}{2\pi}\int \phi \mathrm{d}a =\int \left(\frac{k}{2\pi}\phi \mathrm{d}a+ \phi\delta (p)^\perp\right)~,
\end{equation}
where $\delta(p)^\perp$ is a delta function that restricts the integral to $p$.
Then the equation for $\phi$ implies
\begin{equation}\label{eqn:eoma}
    \mathrm{d}a=-\frac{2\pi}{k}\delta(p)^\perp~.
\end{equation}
If $p$ formed the boundary end points of a curve $\gamma$, then the equation can be solved with $a=-\frac{2\pi}{k}\delta(\gamma)^\perp$. On the other hand, here we only have $U$ inserted at a single point and no such $\gamma$ exists, and the equation cannot be satisfied. Thus, with the appropriate boundary condition $(a = 0)$ the correlation function equals zero despite the fact that the $\mathbb Z_k$ zero-form symmetry is spontaneously broken.

Instead, if we consider the one-point function squared $|\langle U \rangle|^2$, there are two leading contributions: two disjoint disks with $U$ and $U^\dag$ inserted at their boundaries, and a cylinder that connects the two disks (similar to the second line of \eqref{eq:in-out-2} with $\cO_2 = \mathbb 1$).
The first contribution vanishes similarly to the disk one-point function. The second contribution is however nonzero: when normalized in pure BF theory, the correlation function in the topological field theory is $1/k$. More precisely, if the purely gravitational amplitude decreases with the genus and equals $e^{-S_\text{disk}}$ for the disk and $e^{-S_\text{cylinder}}$ for the cylinder, then the leading contribution to the one-point function squared is
\begin{equation}
|\langle U\rangle|^2={e^{2S_\text{disk}-S_\text{cylinder}}\over k}+\cdots~,
\end{equation}
where the $\cdots$ represents sub-leading corrections. 
To see this, note the equation (\ref{eqn:eoma}) can now be solved with $a=-\frac{2\pi}{k}\delta(\ell)^\perp$ with $\ell$ a curve connecting the two insertions of $U$ on the top and the bottom of the cylinder. Then the contribution to the correlation function is nonzero.
Thus, just like in the previous analysis, the $\mathbb{Z}_k$ symmetry is explicitly violated.

\subsection{Differences between global and gauge symmetries} 
\label{sec:global-vs-gauge}

In this section, we will discuss how our argument is affected if the global symmetry is gauged. As previously mentioned, in the holographic context, having a gauge symmetry instead of a global symmetry in the bulk is a common occurrence \cite{Witten:1998qj}.

By gauging a global symmetry $G$, we modify the Lagrangian by typically adding a gauge field, so that it is now invariant under local transformations of $G$.
In the presence of boundaries, we can consider gauge transformations that do not vanish at the boundary but preserve the boundary conditions for the gauge field. In such a case, such a gauge symmetry acts on charged fields which can be placed on the boundary as a global symmetry. 
This means that in the presence of multiple boundaries, after gauging a global symmetry $G$ we find multiple global symmetries each acting separately on their own boundaries.
Going back to our wormhole argument, this means that in order to have a non-vanishing result of the scattering probability, the operator insertions on each boundary need to be invariant under the global transformation, {\it i.e.}~the gauge transformations that preserve the boundary conditions for the gauge field.
That is, if the global symmetry is gauged, the transition probability between two different charge states is zero, and is therefore  not broken by wormholes. Thus, in the example discussed in section \ref{sec:JT+matter}, if we gauge the $U(1)$ symmetry, under which complex scalar $\varphi$ has charge $q$, then $|\<\bar q|q\>|^2 = \< \varphi \varphi\> \<\varphi^* \varphi^*\> = 0$ since the operators inserted on each boundary do not form a gauge singlet, regardless of whether we demand gauge transformations to vanish at each boundary.\footnote{The correlators evaluated here are different than those for the non-local operator, $\varphi e^{\int A} \varphi^*$ which includes a Wilson line stretching between the ends of the wormhole. Rather, here we  only consider insertions of $\varphi$ on the boundary.}

For this $U(1)$ gauge theory, an alternative perspective can be obtained from Gauss's law. 
In such a case, the total charge going though the wormhole geometry needs to vanish due to the equation of motion: $\mathrm{d}*F=j$, where $F$ is the field strength and $j$ is the current.
Integrating this equation over the throat of the wormhole (i.e., over a  closed manifold), we automatically get the constraint $\oint j=0$.

As we shall explain shortly, in a theory with a holographic dual, where the boundary theory is given by an ensemble average, a bulk gauge symmetry means that the symmetry is preserved in each realization of the ensemble. Two symmetries whose ``gauging'' has been extensively discussed in JT gravity \cite{Stanford:2019vob}, and are also present in individual instances of SYK models, are the fermion parity symmetry $(-1)^F$ or the time-reversal symmetry $\cT$ of the boundary dual. We will discuss the role of these symmetries as well as bulk global symmetries in the following subsection.

  \subsection{How the global symmetry $G$ can arise and the factorization problem}   
  \label{sec:how-global-symm-can-arise}
 
 While the previous discussions focused on the gravitational theory, we need to explain how the global symmetry $G$ can arise in the bulk, in a holographic theory with a boundary dual. Since we are considering the contribution of connected geometries for which the bulk partition function does not factorize, we will consider the dual to be given by an ensemble average of boundary theories. If the global symmetry $G$ is present in the bulk, then the same global symmetry $G$ should also be present on the boundary. There are two logical possibilities for the boundary global symmetry $G$: 
\begin{enumerate}[(i)]
     \item  $G$ is a global symmetry for each theory that is part of the ensemble.
     \item $G$ arises only after ensemble averaging and is not a symmetry of each member of the ensemble. 
 \end{enumerate}
 
 We first analyze the case (i).\footnote{Examples of case (i) are JT gravity coupled to 2D gauge theory as discussed in \cite{Iliesiu:2019lfc, Kapec:2019ecr}.} Assuming that $G$ is not spontaneously broken in any of the members of the ensemble, then $|\<\cO_R\>|^2 = :
 \<\cO_R\>\<\cO_R^\dagger\>: = 0$ ($\cO_R$ is some operator charged under the global symmetry $G$ and $:\dots:$ indicates ensemble averaging).  This computation disagrees with the computation in section \ref{sec:replica-wormhole-and-global-symm} of correlators of operators charged under $G$ in the gravitational theory. Therefore, we conclude that such a boundary symmetry cannot correspond to a bulk global symmetry and rather corresponds to a bulk gauge symmetry \cite{Iliesiu:2019lfc, Kapec:2019ecr} which is unaffected by the contribution of the replica wormholes (see \ref{sec:global-vs-gauge}).
 
For case (ii), after ensemble averaging, $:\<\cO_R\>:\,\, =0$; however, there is no reason for the expectation value  $:|\<\cO_R\>|^2:\,\,\, = \,\,\,:\<\cO_R\>\<\cO_R^\dagger\>:$ to vanish once considering the ensemble average. We emphasize this point by considering the example of the SYK model with $N$ Majorana fermions which, after ensemble averaging, has an emergent $O(N)$ symmetry. Such a symmetry acts on the fermionic fields as $\psi_i \to U_{ij} \psi_j$ and on the random coupling as $J_{i_1\dots i_k} \to J_{i_1\dots i_k} (U^{-1})_{i_1 j_1} \dots (U^{-1})_{i_k j_k}$ and leaves the path integral for a single SYK copy invariant:
\be 
\label{eq:SYK-path-integral}
Z_{SYK} \sim \int \mathrm{d}J_{i_1 \dots i_q} \int \mathrm{D}\psi_i \,e^{-\int \mathrm{d}\tau\left[\sum_{i=1}^N\psi_i \partial_\tau \psi_i - (i)^{\frac{q}2} J_{i_1 \dots i_q} \psi_{i_1}\dots \psi_{i_q}\right] \,  }\,,
\ee
where $J_{i_1\dots i_q}$ is the random coupling that is drawn from a Gaussian distribution with variance, $\<j_{i_1\dots i_q}^2\> = J^2 (q-1)!/N^{q-1}$. 
Importantly, since the emergent $O(N)$ symmetry requires that we transform the coupling $J_{i_1\dots i_k}$,  the measure for the integral which averages over this coupling is invariant under these $O(N)$ transformations.

We can now consider a charged operator which transforms under $O(N)$ but is a singlet under the discrete symmetries (such as the fermion parity $(-1)^F$ or the time reversal symmetry $\cT$) that are present in each individual ensemble (we will discuss the case in which the operator is also charged under such discrete symmetries shortly).  We can take such an operator to be $\cO_A = i \psi_{[i} \psi_{j]}$ which transforms in the anti-symmetric representation of $O(N)$, is a singlet under $(-1)^F$, and does not transform under time-reversal when $q \text{ mod } 4 = 2$.\footnote{The factor of $i$ is important in order for $\cO_A$ to be Hermitian. This will be important when discussing how time-reversal acts on $\cO_A$ in SYK models with $q \text{ mod } 4 = 0$. } If we proceed by first integrating-out the coupling $J_{i_1 i_2\dots i_q}$, we can easily show that $\< \cO_A\> = 0$ since there is no spontaneous breaking of the global symmetry in $1D$. However, when studying the correlator which we have primarily discussed in section \ref{sec:replica-wormhole-and-global-symm},  $:|\< \cO_A \>|^2:$, we can no longer use symmetry arguments to show that this correlator vanishes. To obtain $:|\< \cO_A \>|^2:$ we need to consider two copies of the SYK model, coupled through the averaging of the same random coupling: 
\be
\label{eq:exp-value-OA}
:|\< \cO_A \>|^2: \sim \int \mathrm{d}J_{i_1 i_2\dots i_q} &\int \mathrm{D}\psi_i^L \mathrm{D}\psi^R (\psi_{[i}^L \psi_{j]}^L)  (\psi_{[i}^R  \psi_{j]}^R)^\dagger \,\nn \\ &\times 
e^{-\int \mathrm{d}\tau \sum_{P\in \{L, R\}}\left[\sum_{i=1}^N\psi_i^P \partial_\tau \psi_i^P - (i)^{\frac{q}2} J_{i_1 \dots i_q} \psi_{i_1}^P \dots \psi_{i_q}^P \right] \,,  }\,.
\ee
We can now perform the same or different $O(N)$ transformations on the L and R fields. If we perform the same transformation, the operator that we have inserted is invariant and the transformation of the coupling $J_{i_1 i_2\dots i_q}$ remains the same as the one described above; in such a case the path integral in \eqref{eq:exp-value-OA} remains unchanged after the transformation and there is no reason why $:|\< \cO_A \>|^2:$ should vanish. If we perform different transformations on the L and R fields then the path integral over the two SYK copies no longer has an emergent $O(N)$ symmetry since there is no way to act with a unique $O(N)$ transformation on $J_{i_1 i_2\dots i_q}$ which would leave the path integral invariant.  Thus, we find that in contrast to $\< \cO_A \>$, $:|\< \cO_A \>|^2:$ is not protected by any symmetry when considering the ensemble average.

We can also rephrase the above result by coupling the $O(N)$ charge in SYK, $Q^{O(N)}_{ij} = \psi_{[i} \frac{\mathrm{d}}{\mathrm{d}t}\psi_{j]}$ transforming in the anti-symmetric representation of $O(N)$, to a background gauge field $A^{ij}$, {\it i.e.}~by adding $A^{ij} \psi_{[i} \frac{\mathrm{d}}{\mathrm{d}t}\psi_{j]}$ to the Lagrangian in \eqref{eq:SYK-path-integral}. If we consider multiple copies of the system, with a unique random coupling $J_{i_1 i_2\dots i_q} $ as in \eqref{eq:exp-value-OA}, we should also introduce separate copies of the background gauge field (for instance, $A_L$ and $A_R$ when considering two copies). Then the path integral is only invariant under the diagonal gauge transformations that act in the same way on the left and right copies, $A_{L, R} \to h^{-1} A_{L, R} h + h^{-1} \mathrm{d}h $, instead of the most general gauge transformation  $A_{L, R} \to h_{L, R}^{-1} A_{L, R} h_{L, R} + h^{-1}_{L, R} \mathrm{d}h_{L, R} $. This rephrasing emphasizes that the $O(N)$ symmetry, or more generally, any symmetry emergent after ensemble averaging, cannot be dynamically gauged.

We can contrast the above discussion with the case in which we study correlators of an operator that is charged under $O(N)$ but is also charged under some discrete symmetry which is present in each SYK instance. For instance, we can consider $\cO_V = \psi_i$ which is charged under the vector representation of $O(N)$ but is also charged under $(-1)^F$ in any instance of the q-state SYK model.\footnote{In those models, we have that $(-1)^F \cO_V(t) (-1)^F = - \cO_V(t)$.} Alternatively, we can consider the previous operator $\cO_A = i \psi_{[i} \psi_{j]}$ in SYK models with $q \text{ mod } 4=0$, which is odd under time-reversal symmetry.\footnote{In Lorentzian signature, for models with $q \text{ mod } 4=0$, we have that $\cT \psi_i(0) = \psi_i(0)$ from which it follows that $\cT \cO_A(t) \cT^{-1} = -\cO_A(-t)$. A detailed discussion of the action of time-reversal is given in \cite{Stanford:2019vob}. } Since neither time-reversal nor $(-1)^F$ are broken , we conclude that $\< \cO_{A, V}\>=0$ in each instance of the SYK model. As a consequence, there is a similar path integral construction in \eqref{eq:exp-value-OA} that implies $:|\<\cO_{A,V}\>|^2: = 0$. 

Thus, if imagine the bulk dual of the SYK model, $(-1)^F$ and $\cT$ correspond to bulk gauge symmetries and the operators $\cO_{V}$ and $\cO_A$ (for $q \text{ mod } 4 = 0$) are charged under the corresponding bulk gauge fields. The latter corresponds to summing over both orientable and unorientable manifolds, while the former corresponds to summing over spin structures \cite{Stanford:2019vob}.   On the other hand, at the level of the low-energy effective action, the bulk has an $O(N)$ bulk global symmetry whose charge conservation is explicitly violated by the contribution of replica wormholes.\footnote{In the SYK model, there is also a ${1\over N^q}$ violation of $O(N)$ symmetry, in addition to the non-perturbative corrections we are talking about. }

Another useful model which illustrates the phenomenon discussed above is the matrix dual of JT with $\mathbb{Z}_k$ BF theory discussed in section \ref{sec:JT+gauge-theory}.  Due to the bulk gauge $\mathbb{Z}_k$ symmetry, the dual matrix model will have an exact global $\mathbb{Z}_k$ symmetry. This means the Hilbert space can be decomposed into sectors with different representations, which can be labelled by $r\in \mathbb{Z}_k$ \cite{Iliesiu:2019lfc, Kapec:2019ecr}:
\be
\mathcal{H}=\bigoplus_r \mathcal{H}_r~.
\ee
The Hamiltonian will be block diagonal in the basis of different representation $r$:
\be
H=\begin{pmatrix}
H_1 &  &  &\\
 & H_2 & &\\
& & \cdots&\\
& & &H_k
\end{pmatrix}
\ee
where $H_1,\, \dots,\,H_k$ are independent random matrices of the same dimension. 
In this context, the boundary global $\mathbb Z_k$ symmetry, which corresponds to the bulk \textit{gauge symmetry}, is given by $Q=\text{diag}\,(1, \dots, 1, 2, \dots,\, 2, \dots, k, \dots, k)$ for which $[H, Q] = 0$ and the $\mathbb Z_k$ generator is given by $e^{\frac{2\pi i Q}k}$.  The $r$-sector carries charge $r$ under the one-form symmetry that transforms the Wilson line in the corresponding representation.

We now discuss the meaning of the zero-form \textit{bulk global} $\mathbb{Z}_k$ symmetry on the boundary. On the boundary this corresponds to the $\mathbb{Z}_k$ permutation of the different sectors, i.e.~$H_{i} \to H_{(i+1) \text{ mod } k}$, which is an emergent symmetry after taking ensemble average.

It is also useful to take the opposite perspective. Suppose that we have a bulk gravitational theory which has a low-energy effective theory with a global symmetry.\footnote{For example, we can consider this low-energy effective theory to be given by $\cN=8$ supergravity, which has an $SU(8)$ global symmetry. } Then, if we include the contributions of wormholes connecting different boundaries, we will encounter the factorization puzzle discussed in \cite{Maldacena:2004rf,Saad:2019lba}: for instance, the partition function with two boundaries does not factorize. One resolution of this puzzle is that the boundary dual is an ensemble average of theories. In this case, the bulk global symmetry will necessarily arise as an emergent symmetry on the boundary after taking the ensemble average.   If we want to restore factorization in the bulk by finding a UV completion of the theory (in which case we assume that the bulk is dual to a single instance of the ensemble), this inevitably requires an explicit breaking of all bulk global symmetries.

\section{Charged state reconstructions}
  \label{sec:state-reconst}
 
 \subsection{General argument}

  The past computations made it clear that correlators in the ensemble do not obey naive charge conservation properties and, therefore, states with different charges are not necessarily orthogonal. In the following 
%   However, we have not yet made contact with the argument of Banks and Seiberg 
  we will make contact with the argument in \cite{Banks:2010zn}
  regarding the entropy of remnants. 
  Concretely, we would like to understand whether there are indeed a large (or infinite) number of remnant states that are indistinguishable, where the number depends on the dimensions of unitary irreducible representations of $G$. We will show that when including the contribution of all connected geometries to the gravitational path integral, the black hole and remnant states are spanned by a finite, but large, basis of charged states.\footnote{We acknowledge Arvin Shahbazi Moghaddam and Douglas Stanford for useful discussions and suggestions on this topic.   }

For simplicity, we will consider excitations that carry $U(1)$ charges inside the horizon of a thermofield double state. The generalization to the one-sided black hole case\footnote{Such as the black holes that can end on the End-of-World branes.} or to other global symmetries is straightforward.
let us consider a candidate of $K$ charged states $\{|q_1\>,\, \dots\,, \, |q_K\>\}$, created by acting with the operators $\mathcal O_{q_j}$ that carry $U(1)$ charge $q_j$ on the thermofield double state at the middle of the Euclidean evolution.
We would like to show that any other interior state $|\psi\> = \cO_\psi |HH\>$ (see equation \eqref{eq:state-prep}), for some arbitrary operator $ \cO_\psi$, can be reconstructed as a linear superposition of the states $|q_i\>$ for $K> e^{2S_\text{BH}}$. This amounts to showing that $|\psi\>$ can be written as,\footnote{The computation could in principle be generalized to different kinds of orthogonal states in the QFT, not necessarily with different charges. However, when starting with the charged states $|q_i\>$, we do not need to use any dynamical data about higher-point functions in the theory beyond the geodesic approximation. We hope to revisit this calculation in more general QFTs in future work. }
   \be
   \label{eq:q-from-q_i}
   |\psi\> =  \sum_{i=1}^K f^i  |q_i\>,
   \ee
 for some complex coefficients $f^i$.
To simplify our equations a bit in later discussions, we shall use the short-hand notation $|q_f\> \equiv \sum_{i=1}^K f^i |q_i\>$ to represent the state with arbitrary $f^i$.

Our goal is then to maximize the overlap between $|q\>$ and $|q_f\>$:
\be \label{eqn:inner_product}
:\max_{f^i} \frac{\<q_f|\psi\>}{\sqrt{\<\psi|\psi\> \< q_f|q_f\>}}: \,\,\,=\,\,\, :\sum_{i=1}^K (f^{i})^*  \frac{\<q_i|\psi\>}{\sqrt{\<\psi|\psi\> \< q_f|q_f\>}}:\,,
\ee 
where we use the notation $\,\,:\dots:\,\,$ to signal at which point we are taking the ensemble average if preparing the gravitational state in a system which has a boundary dual. 
The fidelity between these two states is equal to the absolute value squared of the inner product and is strictly less or equal to one.
Therefore, if we can show that for any state $|q\>$ the overlap can be arbitrarily close to 1 with some choice of $f^i$, this will mean that the $K$ charged states span a complete basis inside the black hole horizon.

To find the maximum value of the overlap, we can first introduce a Lagrange multiplier $\lambda$ to impose the normalization constraint for $|q_f\>$ and then maximize:
\be
\label{eq:what-to-maximize}
:\max_{f^i,\lambda}\left[  \sum_{i} (f^i)^* \<q_i|\psi\> - \lambda \left(1 - \sum_{i, j} (f^i)^* f^j \<q_i|q_j\>\right) \right]:
\ee   
Extremizing equation \eqref{eq:what-to-maximize} with respect to $f^i$ and $\lambda$ we find: 
\be 
(f)^* M f = 1,\qquad V = \lambda M f\,, \qquad\text{ from which, } \qquad  \lambda = \sqrt{VM^{-1} V}\,, \qquad  f = \frac{M^{-1}V}{\sqrt{V M^{-1} V}}\,, 
\ee
where we have used the simplified vector notation $M_{ij} = \<q_i|q_j\>$ and $V_i = \<\psi|q_i\>$. 
This result implies that the maximum overlap between $|q\>$ and $|q_f\>$ is given by:
\be 
 : \frac{\max_{f^i} \sum_{i} f^i \<\psi|q_i\>}{\sqrt{\<\psi|\psi\>}}: = \frac{
 :\sqrt{V M^{-1} V}:}{\sqrt{\<\psi|\psi\>}} = \frac{:\sqrt{\<\psi|q_i\> \left(\<q_i|q_j\>\right)^{-1} \<q_j|\psi\>}:}{\sqrt{\<\psi|\psi\>}}\,,
\ee
where we have already normalized $|q_f\>$ to have unit norm. 

 To evaluate this on the gravitational side, we will use the replica trick by first considering the quantity $:V M^n V:$ and then analytically continue to $n=-1$. 
 Since we will show that ${:V M^{-1} V:\over \<\psi|\psi\>}$ is arbitrarily close to $1$, it will then follow that ${:\sqrt{V M^{-1} V}:\over \sqrt{\<\psi|\psi\>}}$ is also arbitrarily close to $1$. 
 We would thus like to first evaluate the ensemble average for the function under the square root \be 
 :\<\psi|q_i\> \left(\<q_i|q_j\>\right)^{n} \<q_j|\psi\>:\,\,\, =\,\,\,
 : \sum_{p_{1},\,\dots, \,p_{n+1}\in \{q_1,\,\,\dots\,,\,q_K\} }\<\psi|p_{1}\> \<p_{1}|p_{2}\>\,\dots\,\<p_{n}|p_{n+1}\> \< p_{n+1}|\psi\>:\,,\ee 
 with $n\in \mathbb Z$, using the Euclidean path integral in gravity. 
 This requires us to sum over all possible geometries with $n+2$ asymptotic boundaries, with two charged operator insertions on each boundary, {\it i.e.}~$\cO_{\psi}^\dagger$ and $\cO_{p_1}$ on the first, $\cO_{p_1}^\dagger$ and $\cO_{p_2}$ on the second, and so on, up to   $\cO_{p_{n+1}}^\dagger$ and $\cO_{\psi}$ on the last.
 
   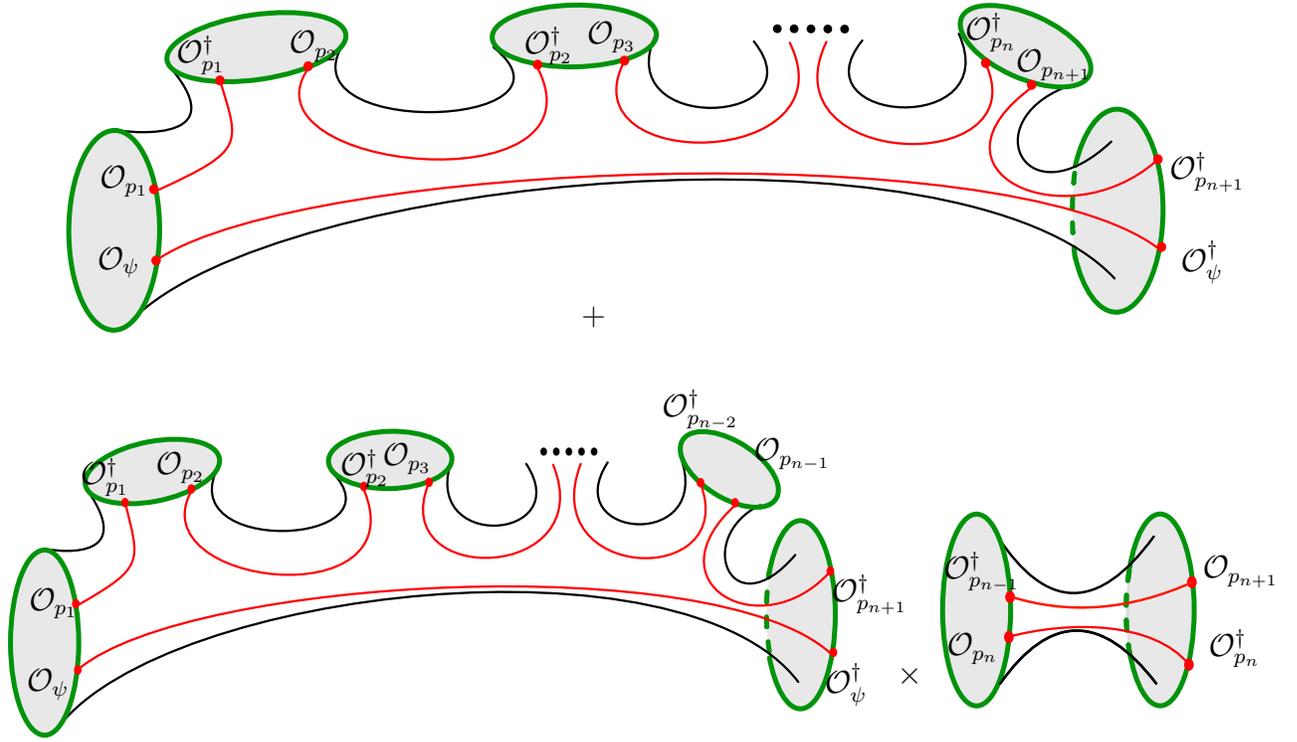
\begin{figure}[t!]
  \begin{center}
 \begin{tikzpicture}[scale=1.2]
 \pgftext{\includegraphics[scale=0.5]{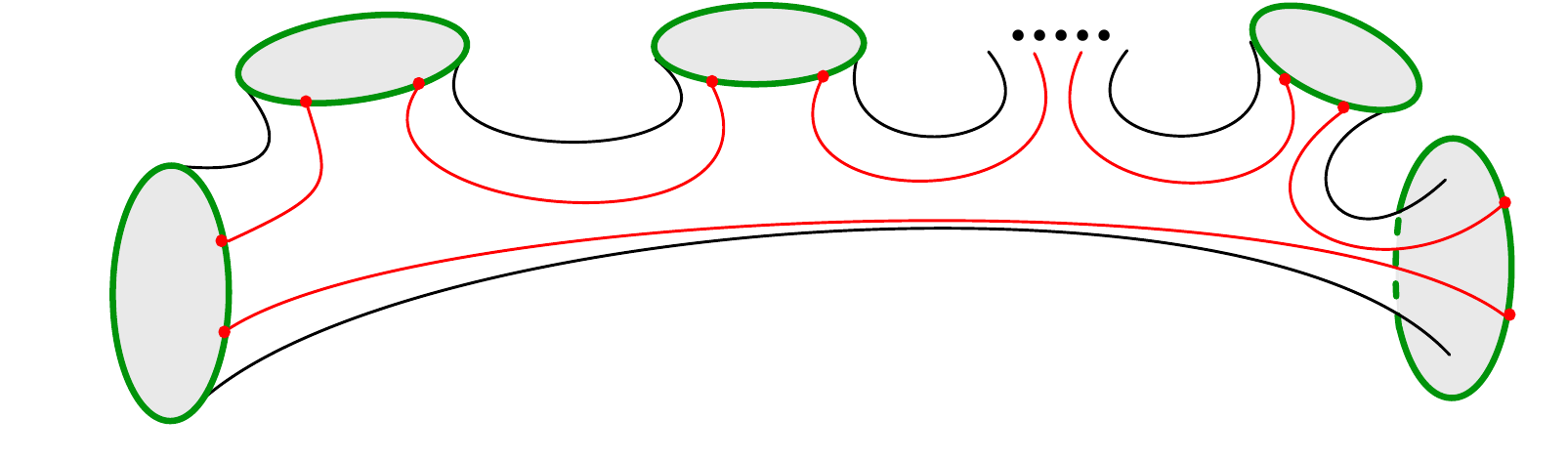}} at (0,0);
 \draw (-5.2,+0.0) node  {$\mathcal O_{{p_1}}$};
  \draw (-5.2,-0.9) node  {$\mathcal O_{\psi\,\,\,}$};
  \draw (-4.35,+1.4) node  {$\mathcal O_{{p_1}}^\dagger$};
   \draw (-3.1,+1.5) node  {$\mathcal O_{{p_2}}$};
 \draw (-0.5,+1.5) node  {$\mathcal O_{{p_2}}^\dagger$};
 \draw (0.2,+1.6) node  {$\mathcal O_{{p_3}}$};    
  \draw (4.4,+1.65) node  {$\mathcal O_{{p_{n}}}^\dagger$};
    \draw (5.1,+1.28) node  {$\mathcal O_{{p_{n+1}}}$};
    \draw (6.8,+0.1) node  {$\mathcal O_{{p_{n+1}}}^\dagger$};
    \draw (6.8,-0.9) node  {$\mathcal O_{\psi\,\,\,}^\dagger$};
    \draw (0,-1.5) node  {$+$};
  \end{tikzpicture}\vspace{0.2cm}\\
   \begin{tikzpicture}[scale=1.2]
 \pgftext{\includegraphics[scale=0.5]{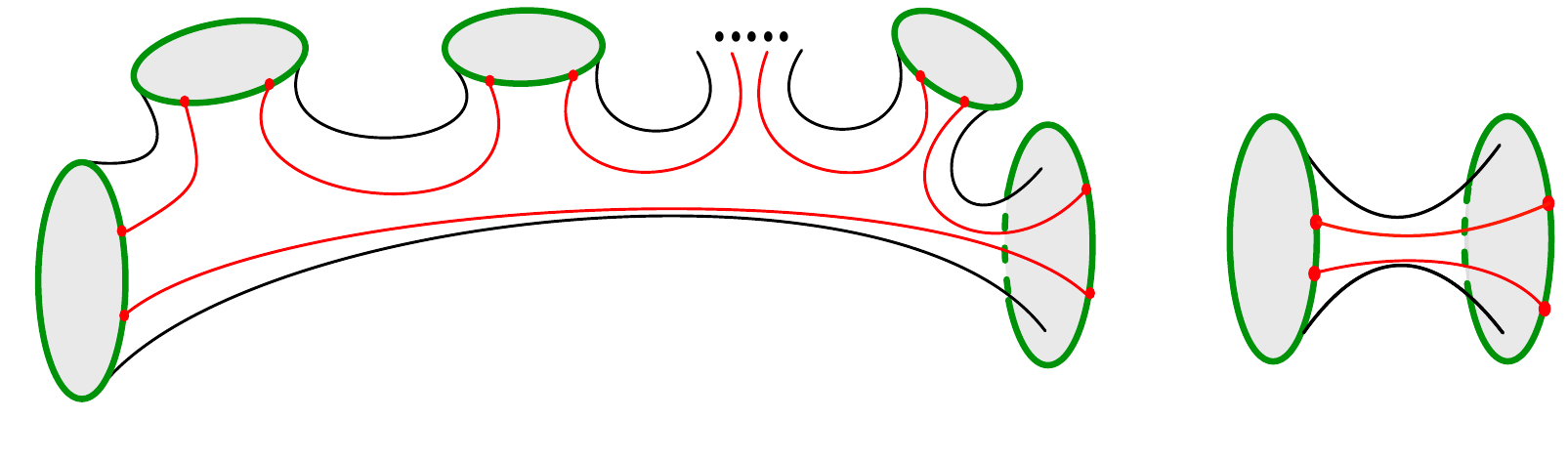}} at (0,0);
  \draw (-2.5+0.67*-5.2,+0.0) node  {$\mathcal O_{{p_1}}$};
  \draw (-2.5+0.67*-5.2,-0.9) node  {$\mathcal O_{\psi\,\,\,}$};
  \draw (-2.5+0.67*-4.35,+1.4) node  {$\mathcal O_{{p_1}}^\dagger$};
   \draw (-2.5+0.67*-3.1,+1.5) node  {$\mathcal O_{{p_2}}$};
 \draw (-2.2+0.67*-0.5,+1.5) node  {$\mathcal O_{{p_2}}^\dagger$};
 \draw (-2.2+0.67*0.2,+1.6) node  {$\mathcal O_{{p_3}}$};    
  \draw (-1.90+0.67*4.6,+2.15) node  {$\mathcal O_{{p_{n-2}}}^\dagger$};
    \draw (-1.35+0.67*5.3,+1.65) node  {$\mathcal O_{{p_{n-1}}}$};
    \draw (-1.5+0.67*6.8,+0.1) node  {$\mathcal O_{{p_{n+1}}}^\dagger$};
    \draw (-1.7+0.67*6.8,-0.9) node  {$\mathcal O_{\psi\,\,\,}^\dagger$};
    \draw (4.3,-0.5) node  {$\mathcal O_{{p_{n\,\,}}\,\,\,}$};
    \draw (4.3,0.35) node  {$\mathcal O_{{p_{n-1}}}^\dagger$};
     \draw (7.1,0.35) node  {$\,\,\,\mathcal O_{{p_{n+1\,}}}$};
        \draw (7.1,-0.5) node  {$\mathcal O_{{p_{n}}}^\dagger$};
            \draw (3.5,-0.8) node  {$\times$};
  \end{tikzpicture}
  
 \caption{\label{fig:all-configs}The first line shows the leading order contributions when evaluation the matrix $M^n$ needed in order to find null states. The second line shows an example of a subleading in $K$ contribution which is only present when $p_{n-1} = p_{n+1}$. When the basis set of states, given by the charges $p_{j} \in \{q_1,\,\,\dots\,,\,q_K\}$, has a dimension $K> e^{2S_\text{BH}}$, the leading contribution is solely given by the first geometry with all other contributions is suppressed in $K$. }
  \end{center}
  \end{figure}
  
  This is essentially the same type of calculation as in the derivation of the Page curve of an evaporating black hole or in the Petz map reconstruction \cite{Almheiri:2019qdq, Penington:2019kki}. In particular, in the limit of $K> e^{2S_\text{BH}}$, the dominating geometry will be the fully connected pinwheel geometry in figure \ref{fig:all-configs}.
 In such a configuration,
every operator $\cO_i$ is connected with $\cO_i^\dagger$ through a bulk propagator.
Summing over all the $i$ indices leads to the maximum power of $K$: 
\be\label{eqn:dominate_saddle}
:V M^n V: \sim \sum_{p_i}\left\< \vspace{1.0cm}\begin{tikzpicture}[baseline={([yshift=-.3ex]current bounding box.center)}, scale=0.6]
 \pgftext{\includegraphics[scale=0.5]{higherbdy1.pdf}} at (0,0);
 \draw (-5.2,+0.0) node  {\tiny{$\mathcal O_{p_{1}}$}};
  \draw (-5.2,-0.9) node  {\tiny{$\mathcal O_{\psi\,\,\,}$}};
  \draw (-4.35,+1.4) node  {\tiny{$\mathcal O_{p_{1}}^\dagger$}};
   \draw (-3.1,+1.5) node  {\tiny{$\mathcal O_{p_{2}}$}};
 \draw (-0.5,+1.5) node  {\tiny{$\mathcal O_{p_{2}}^\dagger$}};
 \draw (0.2,+1.6) node  {\tiny{$\mathcal O_{p_{3}}$}};    
  \draw (4.6,+1.95) node  {\tiny{$\mathcal O_{p_{n}}^\dagger$}};
    \draw (5.9,+1.35) node  {\tiny{$\mathcal O_{p_{{n+1}}}$}};
    \draw (6.8,+0.1) node  {\tiny{$\mathcal O_{p_{{n+1}}}^\dagger$}};
    \draw (6.8,-0.9) node  {\tiny{$\mathcal O_{\psi\,\,\,}^\dagger$}};
  \end{tikzpicture}
  \right\> =K^{n+1} Z_{n+2},
\ee
where $Z_{n+2}$ is the gravitational path integral over the pinwheel geometry with $n+2$ boundaries and insertion of $n+2$ bulk propagators as shown in the figure.

Analytically continuing this result to the limit of $n=-1$, we directly recover the inner product of $\< \psi|\psi\>$:
\be
\label{eq:analytic-cont}
\langle q_f|\psi\rangle=:V M^{-1} V: \sim \left\< \begin{tikzpicture}[baseline={([yshift=-.5ex]current bounding box.center)}, scale=0.6]
 \pgftext{\includegraphics[scale=0.5]{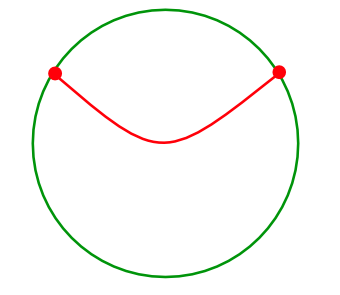}} at (0,-5);
   \draw (-1.65,+1.5) node  {$\mathcal O_{\psi}^\dagger$}; 
 \draw (1.6,+1.5) node  {$\mathcal O_{\psi}$};    
  \end{tikzpicture}
  \right\>  = \<\psi|\psi\>\,,
\ee
which means we have succeeded in reconstructing the state $|\psi\>$ with states $\{|q_i\>\}$.

It is straightforward to see that the above argument continues to hold for more general interior states $|\psi\>$ and $|q_i\>$ beyond theories with global symmetries since the main property we are using is that taking $K$ large prefers the completely connected geometry.
Consequently, this implies that any excitations inside the black hole horizon (including the arbitrary  state $|\psi\>$) can be rewritten as a linear combination of $K$ other states $\{|q_1\>,\, \dots\,, \, |q_K\>\}$ for $K> e^{2S_\text{BH}}$ (or $e^{S_{\text{BH}}}$ for a single-sided black hole).\footnote{Because of the arguments above we can construct the state
\be
|\omega\> = |\psi\>-|q_f\>\,,
\ee
which is null. This is similar in spirit to the existence of null-states recently discussed in the context of $\alpha$-states in \cite{Marolf:2020xie}. However, our interpretation for the Hilbert space of the theory is different than that in  \cite{Marolf:2020xie}. More explicitly, the Hilbert space containing the states in \eqref{eq:q-from-q_i} is not the baby-universe Hilbert space considered in \cite{Marolf:2020xie}; rather, it is the state obtained by acting with charged operators on the gravitational Hartle-Hawking state.
}

We emphasize that this argument applies to gravitational theories in general spacetime dimensions, with the two-dimensional figures replaced by the higher-dimensional analogue.
For sufficiently large number of states in a basis whose dimension is greater  than $e^{2S_\text{BH}}$ (or $e^{S_\text{BH}}$ for a single-sided black hole), the connected geometry dominates over all other in the Euclidean path integral, and we observe that the basis is in fact over-complete and does not correspond to distinct superselection sectors.
This is sufficient if we wish to find the analytic continuation of $:V M^{-1} V:$ from \eqref{eq:analytic-cont}. In the next subsection we will analyze possible corrections to \eqref{eq:analytic-cont}.

With this new understanding of the over-completeness of charged states in gravitational theories, in the next section we will re-analyze the fate of remnants in the argument by Banks and Seiberg, reviewed in section \ref{sec:banks-seiberg-arg}.

\subsection{Planar Resummations in JT gravity}
\label{sec:planar-resumm}
In this section, we discuss the state reconstruction in JT gravity.
First, we notice that as long as $K\gg 1$ and $e^{S_0}\gg 1$, the dominating contribution to $:V M^n V:$ are the planar geometries just like in the case of the end of the world (EoW) brane model, studied in \cite{Penington:2019kki} (see also \cite{Dong:2020iod,Marolf:2020vsi, Akers:2020pmf}).
Therefore, we can use the same resolvent technique to solve this exactly.
The strategy will be to first consider insertion of the resolvent operator $\bold{R}$ of $M$:
\be
: V \bold{R}(\lambda) V:\,\,\,\equiv\,\,\,:V{1\over \lambda-M}V:,~~~:V M^n V:\,\,\,={1\over 2\pi i}\oint \mathrm{d}\lambda \lambda^n : V \bold{R}(\lambda) V:\,,
\ee
and then do an analytic continuation to $n=-1$. Due to the typical branch cut structure of the resolvent, the integration contour needs to be deformed and we will see how that precisely works.
The boundary condition of $: V\bold{R}(\lambda) V :$ corresponds to an infinite summation of an indefinite number of boundary circles:
\be
 \begin{tikzpicture}[baseline={([yshift=-.5ex]current bounding box.center)}, scale=0.6]
 \pgftext{\includegraphics[scale=0.9]{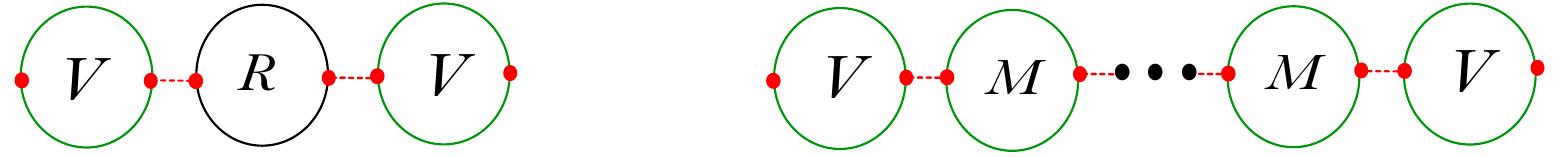}} at (0,0);
   \draw (-2.10,0) node  {{\Large$= \sum$}}; 
  \end{tikzpicture}\,.
\ee
The corresponding bulk geometries can be classified by the geometry $\mathcal{M}$ that connects the two $V$ type boundaries, which must exist for the answer to be nonzero.
To specify $\mathcal{M}$, we also need to know how $\mathcal{M}$ ends on the various $M$ type boundaries and we will use the notation $\mathcal{M}_{2,l}$ to represent the geometry that ends on $l$ such $M$ type boundaries.
Together with the two $V$ type boundaries, $\mathcal{M}_{2,l}$ has a total of $l+2$ boundaries.
Between these boundaries we still have an infinite sum of planar geometries which can be rewritten as an insertion of a resolvent:
\be 
 \begin{tikzpicture}[baseline={([yshift=-.5ex]current bounding box.center)}, scale=0.6]
 \pgftext{\includegraphics[scale=0.7]{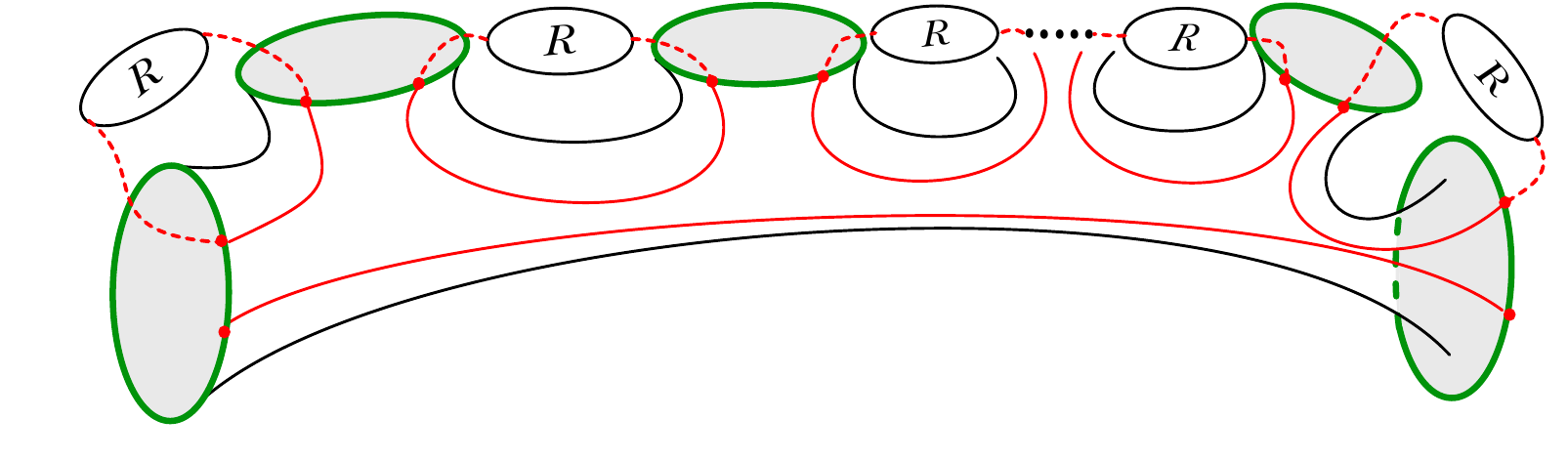}} at (0,0);
    \draw (-10,0) node  {\Large{$\sum_{\mathcal{M}_{2,l}}$}}; 
      \draw (-7.5,-1) node  {{$\cO_\psi$}}; 
            \draw (9.5,-1) node  {{$\cO_\psi^\dagger$}}; 
  \end{tikzpicture}
\ee 

Summing over all the $\mathcal{M}_{2,l}$s, we get an exact expression of $:V \bold{R}(\lambda) V:$ in terms of the resolvent $R(\lambda)\equiv\Tr \bold{R}(\lambda)$:
\be\label{eqn:geometric_series}
:V \bold{R}(\lambda) V:=\sum_{l=0}^{\infty}  R^{l+1} (\lambda)\, Z_{l+2}
\ee
where we used the notation $Z_{l+2}$ to represent the gravitational path integral over the geometry $M_{2,l}$ (this is the same notation as in equation \ref{eqn:dominate_saddle}).
The gravitational path integral $Z_{l+2}$ in JT can be derived using the same cutting and gluing procedure we used in section \ref{sec:JT+matter}.
By cutting along the propagators, we can separate the pinwheel geometry into two pieces, each containing $l+2$ numbers of the geodesics $\ell_i$ and the semi-boundaries $\beta_i={\beta\over 2}$.
Since there is no additional operator insertion in such geometry, all the boundaries have the same energy and the full gravitational path integral can be written as:
\be
\int \mathrm{d}E \rho(E) \prod_i e^{-\beta_i E} \langle E|\ell_i\rangle\,.
\ee
Gluing two copies of this geometry together with the weighting of the propagators $e^{-m \ell_i}$, we get:
\be
\begin{split}\label{eqn:Zn}
Z_{n}&=e^{S_0 (2-n)} \int \mathrm{d}E_1 \mathrm{d}E_2 \rho(E_1)\rho(E_2) e^{-{n\over 2}\beta (E_1+E_2)} \prod_i\int \mathrm{d}\ell_i \langle E_1|\ell_i\rangle e^{-m\ell_i}\langle \ell_i|E_2\rangle\\
&= \int \mathrm{d}E_1 \mathrm{d}E_2 e^{2S_0} \rho(E_1)\rho(E_2) y_{E_1,E_2}^{n};~~~y_{E_1,E_2}=e^{-S_0}e^{-{1\over 2}\beta(E_1+E_2)}\langle E_1|\mathcal{O}^{\dagger}\mathcal{O}|E_2\rangle\,.
\end{split}
\ee
We can understand this formula as a Boltzmann summation over the product of $l+2$ correlators in the energy basis $E_{1,2}$, and we expect that this is a general result, that holds beyond JT gravity.
On the other hand, if we go to the microcanonical ensemble rather than the canonical ensemble, the result of the path integral will just be the integrand.
Plugging this into equation \eqref{eqn:geometric_series}, and summing over the geometric series, one gets:
\be\label{eqn:VRV1}
\begin{split}
:V\bold{R}(\lambda) V:&=e^{2S_0}\int \mathrm{d}E_1 \mathrm{d}E_2 \rho(E_1)\rho(E_2) {y^2 R \over 1- y R };\\
\langle q_f|\psi\rangle=:VM^{-1}V:&=e^{2S_0}\int \mathrm{d}E_1 \mathrm{d}E_2 \rho(E_1)\rho(E_2) y \oint{\mathrm{d}\lambda\over 2\pi i\lambda} {y R \over 1- y R }.
\end{split}
\ee
Therefore, if we know the value of the resolvent $R$, we know the full answer of $:V\bold{R} V:$ and also $:VM^{-1}V:$.
Using free probability theory, or by classifying the planar diagrams as what we did before, one finds that the resolvent satisfies a Schwinger-Dyson (SD) equation (\cite{Penington:2019kki}):
 \be\label{eqn:SD equation}
 \lambda R=K+\sum_{n=1} Z_n R^n=K+\int \mathrm{d}E_1 \mathrm{d}E_2 e^{2S_0}\rho(E_1)\rho(E_2){y R\over 1-y R}.
 \ee
 
The physics of this Schwinger-Dyson equation is quite rich, and it includes the phase transition between different Renyi entropies of the system; generally there is no known exact method of solving this equation, apart from numerics \cite{Penington:2019kki}.
Below we consider a simpler equation by directly going to the microcanonical ensemble and then draw some general lessons from that. We also present the computation in the canonical ensemble in appendix \ref{sec:canonical-ensemble}.

Going to the microcanoincal ensemble, we are fixing the energy of $E_{1,2}$ to be in a small energy window $(E,E+\delta E)$.  This leads to a simplified version of equation (\ref{eqn:VRV1}) and equation (\ref{eqn:SD equation}):
\begin{eqnarray}
:V\bold{R}(\lambda)V:=e^{2\bold{S}}{y^2 R\over 1-y R}\,,\label{eqn:VRVMicro}\\
\lambda R=K+e^{2\bold{S}}{yR\over 1-y R}\,, \label{eqn:SDMicro}
\end{eqnarray}
where we used the notation $e^{\bold{S}}\equiv \delta E e^{S_0} \rho(E)$ and $Z_n=e^{2\bold{S}}y^n$.
Solving the quadratic equation \eqref{eqn:SDMicro}, we get:
\be
\begin{split}
    R(\lambda)&={1\over 2y}+{K-e^{2\bold{S}}\over 2\lambda}-{1\over 2\lambda y}\sqrt{(\lambda-\lambda_+)(\lambda-\lambda_-)};~~~\lambda_{\pm}=y(e^{\bold{S}}\pm \sqrt{K})^2\,,\\
D(\lambda)&={1\over 2\lambda y}\sqrt{(\lambda_+-\lambda)(\lambda-\lambda_-)}+\delta(\lambda)(K-e^{2\bold{S}})\theta(K-e^{2\bold{S}})\,,
\end{split}
\ee
where $D(\lambda)\equiv{1\over 2\pi i}(R(\lambda-i\epsilon)-R(\lambda+i\epsilon))$ is the density of states.
We see that when $K< e^{\bold{S}}$, the density of state is fully support in $(\lambda_-,\lambda_+)$.
After $K>e^{\bold{S}}$, the support of $D(\lambda)$ splits into two parts: there are $K-e^{2\bold{S}}$ states located at $\lambda=0$ and $e^{2\bold{S}}$ distributed between $\lambda_-$ and $\lambda_+$.

Finally, let us look at the inner product $\langle q_f|\psi\rangle$.
Combining equation \eqref{eqn:VRVMicro} with equation \eqref{eqn:SDMicro} we have:
\be
:V\bold{R}(\lambda)V:=(\lambda R -K ) y\,,
\ee
which has a branch cut coming from the resolvent.
This gives us:
\be
:VM^{-1} V:=\lim_{n\rightarrow -1} \oint \mathrm{d}\lambda \lambda^{n} V\bold{R}(\lambda)V=\lim_{n\rightarrow -1} {1\over 2\pi i}\oint \mathrm{d}\lambda \lambda^{n}(\lambda R(\lambda)-K) y\,.
\ee
When we do the analytic continuation in $n$, $\lambda^{n+1}$ will generically have a branch cut from $0$ to infinity.
To avoid this issue, we can first deform the integration contour of $\lambda$ to go around the branch cut of $R(\lambda)$ and then analytically continue in $n$.
This leads to the final expression:
\be
\langle q_f|\psi\rangle=:VM^{-1} V:=y\int_{\lambda_{-}}^{\lambda_+}\mathrm{d}\lambda D(\lambda)= \begin{cases}K y~~~~~K< e^{2\bold{S}}\\
e^{2\bold{S}}y~~~~K\geq e^{2\bold{S}}\end{cases}.
\ee
Recall that since $\<\psi|\psi\>=Z_1=e^{2\bold{S}}y$, we see that the inner product between $|\psi\>$ and $|q_f\>$ is equal to one when $K$ is bigger than $e^{2\bold{S}}$.
This means in the microcanoincal ensemble case that any state $|\psi\>$ can be reconstructed from the $K=e^{2\bold{S}}$ charge states $\{|q_1\>,...|q_K\>\}$.
In other words, they form a complete basis of the states in the microcanonical black hole.

\section{The fate of remnants}
\label{sec:BH-remn}

    In section \ref{sec:state-reconst}, we argued that states that are linearly independent in a quantum field theory can become over-complete when this theory is coupled to gravity.
    In particular, we showed that when coupling a theory with global symmetry to gravity, then a generic excitation inside of a black hole $|\psi\>$ can be written as a linear combination of a basis of $K$ charged states,  $\{|q_1\>, \, \dots, |q_K\>\}$, when $K$ is larger than $e^{2S_\text{BH}}$ (or $e^{S_\text{BH}}$ for a single sided black hole). 
    While the explicit computation in section \ref{sec:state-reconst}, was primarily done for charged excitations in the thermal-field double state, a similar computation should apply to the states inside of an evaporating black hole, which can be modeled by coupling the system shown in figure \ref{fig:all-configs} to a bath, represented by a large region in flat-space, where gravitational effects can be ignored (for example, see \cite{Almheiri:2019psf}). 
    
    \begin{figure}[t!]
    \centering
     \begin{tikzpicture}[baseline={([yshift=-.5ex]current bounding box.center)}, scale=0.6]
 \pgftext{\includegraphics[scale=0.5]{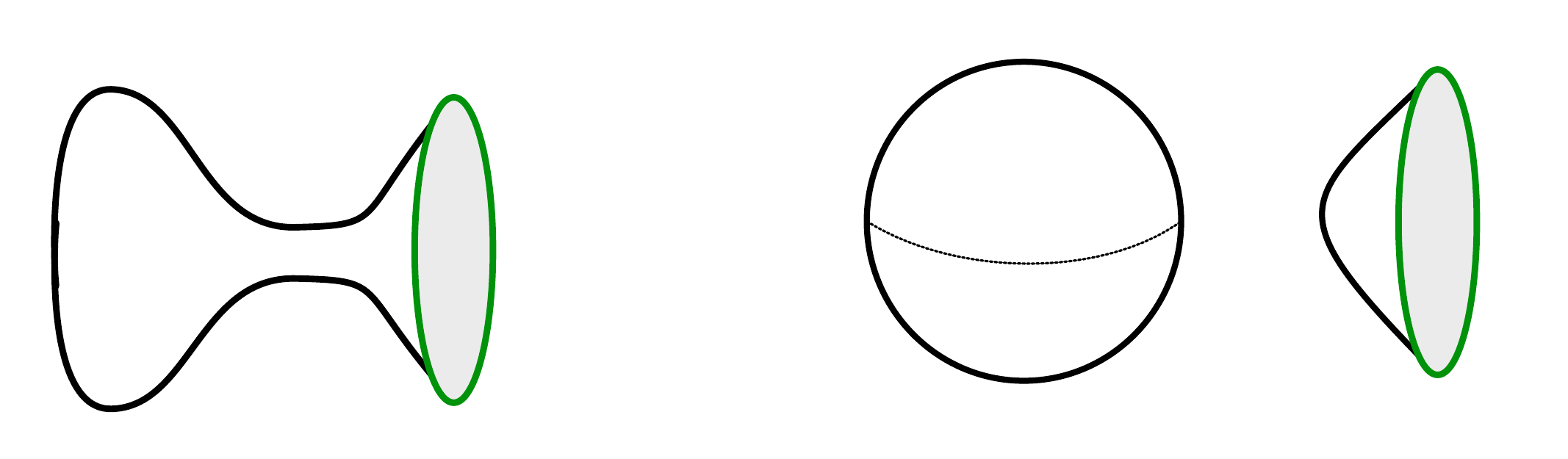}} at (0,0);
   \draw (-1,0) node  {{\Large$\longrightarrow$}}; 
   \draw (-5.5,-2.7) node  {{Black hole at late times}}; 
   \draw (2.8,-2.7) node  {{Closed universe}}; 
  \end{tikzpicture}
    \caption{Spatial section of black hole at late times, decaying into a closed universe.}
    \label{fig:closed_univ}
\end{figure}

    As in \cite{Banks:2010zn}, we can imagine that the gravitational effective theory which we use in the path integral computation can be trusted up to the point where a remnant is formed.
    That is the radius of the black hole horizon is given by $r_h = X L_\text{Pl}$ and its Bekenstein-Hawking entropy is given by $e^{S_\text{BH}} = e^{\pi X^2}$. 
    In the argument made in \cite{Banks:2010zn}, each remnant state that carried the representation $R$  of the global symmetry was considered to be indistinguishable and independent. 
    This leads to a large degeneracy of remnant states which, with some technical assumptions mentioned in section \ref{sec:banks-seiberg-arg}, contradicted the covariant entropy bound, or its more refined version: the ``central dogma'' that a black hole describes a quantum mechanics system with $S_\text{BH}$ degrees of freedom. 
   However, because we have found that by considering the contribution of replica wormholes, the Hilbert space in the gravitational theory is over-complete, the situation is now much less dramatic.
   The complete basis of the states inside of the remnant can be chosen to be the $K$ charged states.
   Since $K$ is of order $e^{S_\text{BH}}$, this means that the degree of freedom  of the remnant is given by  $ \log K \sim \pi X^2$ which is consistent with the ``central dogma''.\footnote{Notice that there could be order one corrections to the Bekenstein-Hawking entropy. For instance, it is well-known that the black hole entropy receives $\sim \log r_h^2$ corrections with an unfixed sign.}

We can now speculate about the ultimate fate of these remnants. Historically, remnants were argued to be non-existent due to the thermodynamic instability caused by the large internal entropy \cite{Giddings:1993km,Susskind:1995da}.
From our discussion, the possible remnants forming from black hole evaporation do not encounter such issues since the entropy still obeys the Benkenstein-Hawking bound and can, in principle, exist in nature. However, it might no longer be appropriate to call such an object a remnant since most information of the black hole has ``escaped'' through the replica wormhole and is encoded in the Hawking radiation. For instance, information about the global symmetry charge of the initial black hole is no longer captured by the remnant states. Because of that, if such objects exist at the end of evaporation, perhaps a better name for them is ``faint remnants''.

We also see that whether or not the remnant is formed from global symmetry charged states does not make a difference due to the large violation of the global symmetry inside the remnant. 
For an outside observer, the remnant seems to be nothing more special than an ordinary small black hole and it seems reasonable that it will eventually decay into a closed universe including the whole interior of the black hole, as portrayed in figure \ref{fig:closed_univ}.

\begin{figure}[t!]
    \centering
    \includegraphics[scale=0.5]{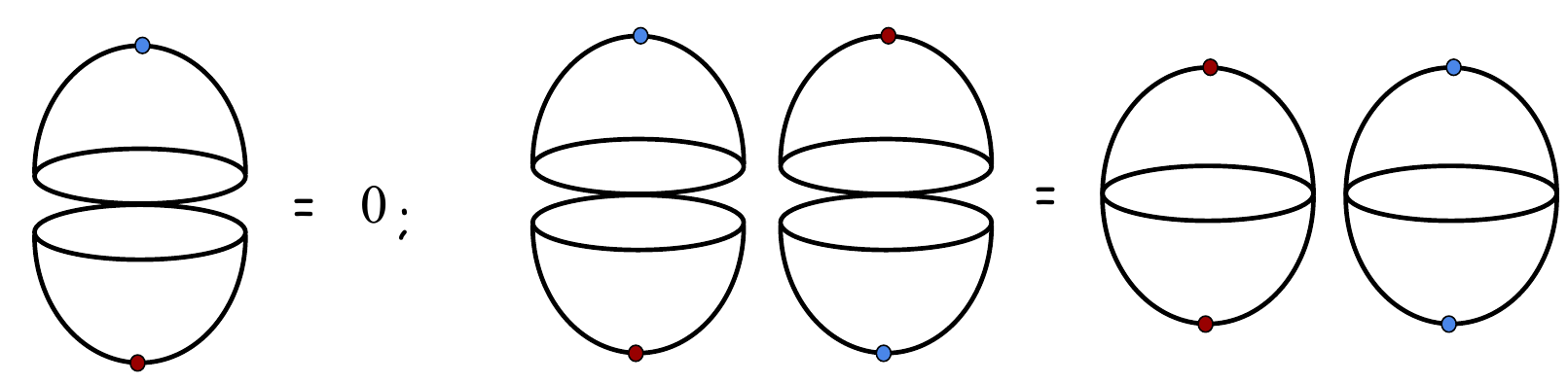}
    \caption{The inner product between different states in a closed universe. The brown and blue dots represent insertion of operators with different charges. On the left figure, the inner product between these two states is zero. On the right figure, the fidelity of these two states is equal to one, which means the two states are equivalent up to a phase.}
    \label{fig:CUinnerproduct}
\end{figure}

Extrapolating our result in section \ref{sec:state-reconst} to the closed universe case, we conclude that all the different states inside the closed universe are equivalent, up to a random phase:
\be\label{eqn: BUstates}
|q_i\rangle_{\text{CU}}= e^{i\theta_i}|\text{CU}\>,
\ee
This is consistent with the following gravitational picture shown in figure \ref{fig:CUinnerproduct}:
the states $|q_i\>$ and $|q_j\>$ in a closed universe can be prepared by inserting different operators in the past. 
If we calculate their inner product $\<q_i|q_j\>$ directly in the bulk, the answer will be zero. 
However, the fidelity between these two states $|\<q_i|q_j\>|^2$ is equal to one since the closed universe can just connect between the two copies.
This is a strong hint that $|q_i\>$ and $|q_j\>$ are actually the same state up to a  phase which is random, so that after ensemble average the inner product vanishes.
In other words, our computation suggests that \emph{the Hilbert space of a closed universe is one-dimensional.}

Using this property, let us reexamine Hawking's original argument (section \ref{sec:hawking-arg}) about the charge violation that can occur during black hole evaporation.
If the initial state of the universe was in a singlet state, after creating a black hole and letting it evaporate, the rest of the universe will be in a charge singlet state together with the baby universe.
Using the relation between the different charged states inside of the closed universe (as in \eqref{eqn: BUstates}), we find that the final state will become a random superposition of charged states.
Schematically we have:
\be
|\text{final state}\>=\sum_i |q_i\>_{\text{CU}}|\bar q_i\>_{\text{rest}}=\sum_{i} e^{i\theta_i} |\bar q_i\>_{\text{rest}} |\text{CU}\>
\ee
Once again, to make a connection with Hawking's result, we can consider the density matrix of the final state and perform an ensemble average. This gives us a thermal distribution of the charge sectors due to the randomness of the phases $\theta_i$.

\section{Conclusion and comparison to past arguments}
\label{sec:conclusion}

This work provides a new argument about the violation of exact global symmetries in quantum gravity, using replica wormholes.
We argue that the existence of replica wormholes predicts a non-vanishing transition probability (or inner product) between different charged states. This is a non-perturbative violation of the global symmetry charge conservation, and such a violation holds even in a theory with no explicit breaking of the global symmetry up to arbitrarily high energy scales.

The main mechanism for this violation stems from the fact that the transition probability involves computations in two copies of the system. The operator insertions can form a singlet on the replica wormhole which connects the two copies, while the leading disconnected contribution vanishes due to the presence of the global symmetry.
As a consequence, quantum gravity does not allow super-selection sectors coming from any exact global symmetries. In principle, our computation in the toy model of JT gravity coupled to matter with a $U(1)$ charge can be embedded to describe the scattering probability from baryons to leptons for higher-dimensional near-extremal Reissner-Nordstr\"om black holes, coupled to the Standard Model or to some of its extensions that preserve the baryon-lepton number global symmetry.\footnote{Generically, the contribution of higher topology or multi-boundary geometries to the near-horizon path integral is untrustworthy \cite{Iliesiu:2020qvm}. That is because there are numerous corrections which will kick-in (such as corrections to the dilaton potential or from Kaluza-Klein modes \cite{ Nayak:2018qej, Moitra:2018jqs, Iliesiu:2020qvm}), before the $O(e^{-S_\text{BH}})$ corrections come in from higher topologies. However, because the replica wormhole is the leading non-vanishing contribution due to the $U(1)$ global symmetry, the calculation of the scattering probability is now trustworthy.}  
In contrast, if we gauge the global symmetry, then the transition probability between charged states is still zero, even when including the contribution of replica wormholes.
The reason for this is that, after we gauge the symmetry, there will be individual global symmetries associated with each of the boundaries which provide a stronger constraint on correlation functions. 
An alternative point of view is that there will be a constraint by Gauss's Law on the throat of the wormhole after we couple to a gauge field.

We have also argued that the states that are orthogonal in an ordinary quantum field theory can form an over-complete basis when coupling that quantum field theory to gravity.
In the context of black holes and their late-time remnants, this leads to a verification of the "central dogma" that such systems have $S_\text{BH}={A\over 4 G_{\text N}}$ degrees of freedom, even when the number of distinct unitary irreducible representations of the global symmetry is large, or infinite.
In addition, our calculation suggests that the Hilbert space of a closed universe is one dimensional.

Let us compare the results of this paper to the previous arguments of \cite{Hawking:1982dj, Banks:2010zn,Harlow:2018tng, Harlow:2020bee} against global symmetries in quantum gravity (reviewed in section \ref{sec:review-past}).  Due to the contribution of replica wormholes, our result provides a concrete base for Hawking's intuition that global symmetry charge is not conserved during the process of black hole evaporation, and our computation extends his argument to eternal black holes.
Regarding the dollar matrix discussion in section \ref{sec:review-past}, our gravitational computation predicts a non-factorization property of the dollar matrix coming from the ensemble average:
\be
:\$_{m m';n n'}:\,\,\,= \,\,\,:{S_{m n} S^*_{m' n'}}:\,\,\,\neq \,\,\, :{S_{m n}}:~:{S^*_{m' n'}}:\,,
\ee
where $:\dots:$ denotes the point at which we consider an ensemble average. 

We also see that at late times black holes in theories with a global symmetry have a much smaller degeneracy,  $\sim e^{A\over 4 G_{\text N}}$, than naively expected in the absence of replica wormholes \cite{Banks:2010zn}. Our results imply that there is no apparent contradiction in the existence of black hole remnants, which, due to the lack of the global symmetry charge conservation, could, in principle, fully evaporate. 

Finally, in arguing for the violation of global symmetries in quantum gravity, we have not used the extremal island formula as in \cite{Harlow:2020bee} and rather, we directly used the contribution of replica wormholes.
 In particular, our computation quantifies how such geometries lead to the violation of the global charge conservation and makes direct contact with the possible fate of remnants.  Finally, while previous arguments suggested that global symmetries are absent in the effective field theory beyond some energy level, we are able to explain what the presence of the global symmetries in the effective gravitational description implies for a holographic theory. We will expand on this below.

Throughout the paper, when the gravitational theory is holographic, we interpret the boundary theory as an ensemble average. From that point of view, the global symmetry in the bulk is a result of ensemble averaging of the boundary theories, where, in each member of the ensemble, the global symmetry is absent.
An analog of this situation is the $O(N)$ symmetry of the SYK model. For each realization, the coupling constant breaks the global symmetry explicitly, but the distribution of the coupling constant is $O(N)$ invariant. This leads to the existence of an emergent $O(N)$ global symmetry after ensemble averaging.
Gauge symmetries in the bulk, on the other hand, are boundary global symmetries that are present in all members of the ensemble, and, in this case, the contribution of replica wormholes does not lead to a violation of charge conservation.

Our wormhole calculation indicates an explicit breaking of the global symmetries in the effective Lagrangian of a single instance of the ensemble. 
This breaking suggests that any UV completion of this effective Lagrangian, which resolves the factorization puzzle, should not have any manifest global symmetries; rather, global symmetries in quantum gravity are a manifestation of ensemble averaging and absence of factorization.

\section*{Acknowledgement}
We thank Juan Maldacena, Arvin Shahbazi Moghaddam, Stephen Shenker and Douglas Stanford for comments on a draft and for helpful discussions. We thank 
Edgar Shaghoulian for comments on a draft. The work of P.-S.\ H.\ is supported by the U.S. Department of Energy, Office of Science, Office of High Energy Physics, under Award Number DE-SC0011632, and by the Simons Foundation through the Simons Investigator Award. LVI was supported in part by the Simons Collaboration on Ultra-Quantum Matter, a Simons Foundation Grant with No. 651440. 
ZY is supported in part by the Simons Foundation through the It from Qubit Collaboration.

\appendix

\section{Squared one- and two-point function}
\label{app:one and two-point function}
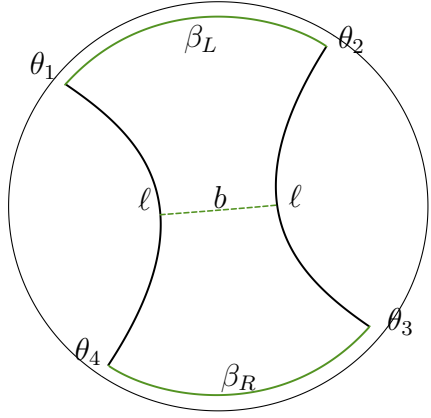
\begin{figure}[h!]
\begin{center}
 \begin{tikzpicture}[scale=0.5]
 \pgftext{\includegraphics[scale=0.4]{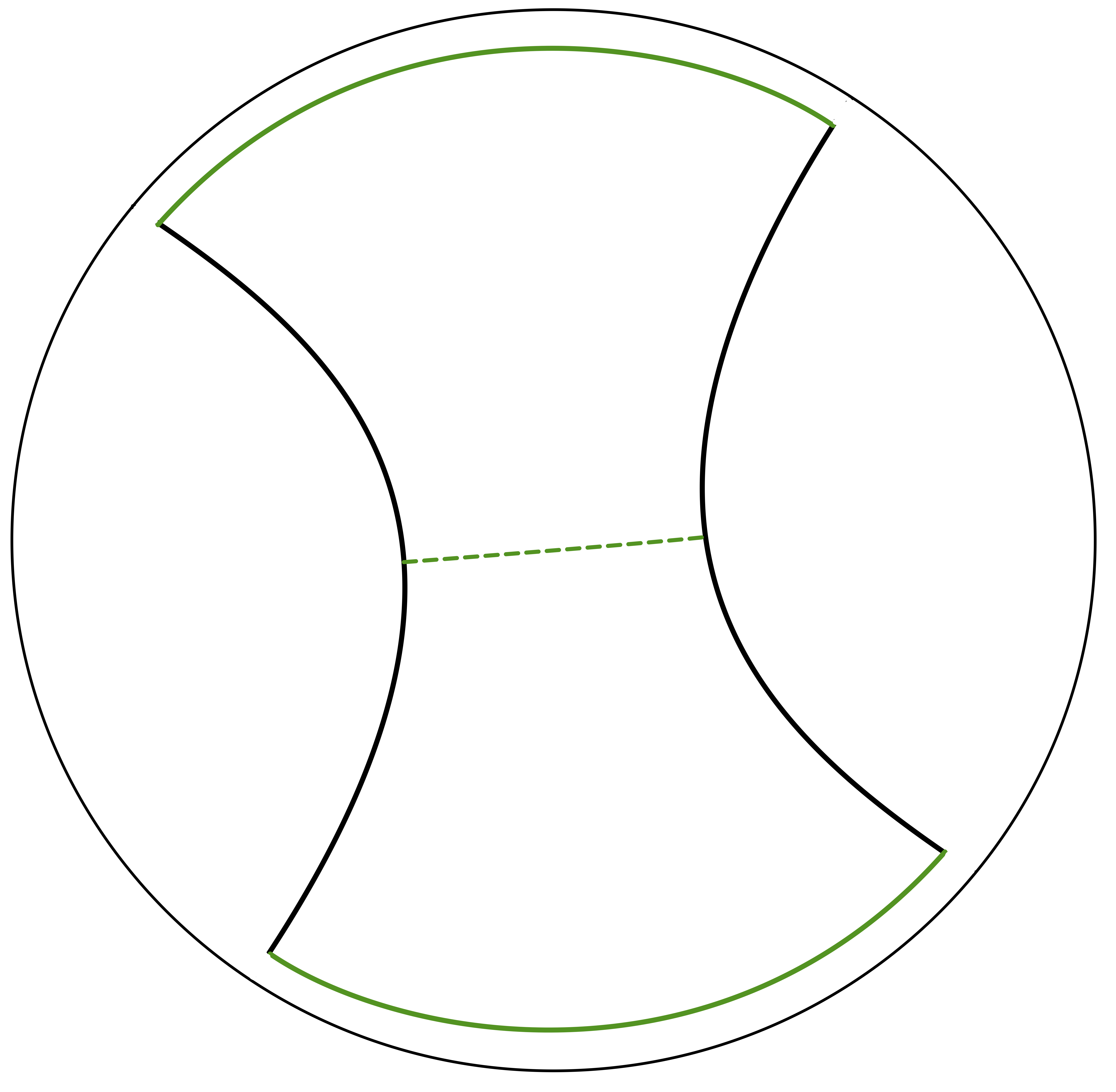}} at (0,0);
  \draw (0,+0.3) node  {$b$};
  \draw (-0.5,+4.5) node  {$\beta_L$};
 \draw (0.5,-4.5) node  {$\beta_R$};
  \draw (-2,+0.25) node  {$\ell$};
   \draw (2,+0.35) node  {$\ell$};
   \draw (-4.7,+3.8) node  {$\theta_1$};
   \draw (3.5,+4.5) node  {$\theta_2$};
      \draw (-3.5,-3.8) node  {$\theta_4$};
   \draw (4.8,-3.0) node  {$\theta_3$};
   
\end{tikzpicture}
\caption{Construction of the wormhole geometry on the Poincar\'e disk.}
\label{fig:hyperdisk}
\end{center}
\end{figure}

In this appendix, we review the calculation of the squared one-point  and two-point functions in JT gravity \cite{Saad:2019pqd,Stanford:2020wkf}.
The basic ingredient is the matrix element of the evolution of the two-sided Hamiltonian $e^{-\beta_L H_L-\beta_R H_R}$ in the geodesic basis:
\be
\langle \ell|e^{-\beta_L H_L-\beta_R H_R}|\ell'\rangle=\int \mathrm{d}E \rho(E) e^{-(\beta_L+\beta_R)E}\langle \ell|E\rangle\langle E|\ell'\rangle\,,
\ee
where $\rho(E)={1\over 2\pi^2}\sinh(2\pi\sqrt{2E})$ is the density of states and $\langle \ell|E\rangle=4 K_{2i\sqrt{2E}}(4 e^{-\ell/2})$.
This can be understood as the gravitational path integral over a rectangular geometry with two geodesic boundaries $\ell,\ell'$ and two asymptotic boundaries (figure \ref{fig:hyperdisk}).
The other ingredient is the geodesic approximation of the two-point function which is simply given by
\be
\langle \ell |\mathcal{O}_q\mathcal{O}_{-q}|\ell'\rangle=e^{-m\ell}\langle \ell|\ell'\rangle\,.
\ee
This directly gives us  the absolute value of the squared one-point function:
\be
|\langle \mathcal{O}_q\rangle|^2=\int \mathrm{d}\ell e^{-m\ell}\langle \ell|e^{-\beta_L H_L-\beta_R H_R}|\ell\rangle=\int \mathrm{d}E \rho(E) e^{-(\beta_L+\beta_R)E} \langle E|\mathcal{O}_q\mathcal{O}_{-q}|E\rangle
\ee
where $\langle E|\mathcal{O}_q\mathcal{O}_{-q}|E'\rangle=\int d\ell \langle \ell|E\rangle\langle E'|\ell\rangle e^{-m\ell}$ is the two-point function in the energy basis:
\be
\langle E|\mathcal{O}_q\mathcal{O}_{-q}|E'\rangle={\Gamma(m\pm i(\sqrt{2E}\pm \sqrt{2E'}))\over 2^{2m+1}\Gamma(2m)}
\ee
where the $\pm$ sign means we need to take product of all the four gamma functions coming from different choice of the $\pm$ signs.

Similarly, the two-point function squared can be calculated from gluing two rectangular regions along the two geodesics where the operators are inserted:
\be
\begin{split}
|\langle \mathcal{O}_{q_1}\mathcal{O}_{q_2}\rangle|^2&=\int \mathrm{d}\ell \mathrm{d}\ell' e^{-m_1\ell}\langle \ell|e^{-\beta_{L1} H_L-\beta_{R1} H_R}|\ell'\rangle e^{-m_2\ell'} \langle \ell'|e^{-\beta_{L2} H_L-\beta_{R2} H_R}|\ell\rangle\\
&=\int \mathrm{d}E_1 \mathrm{d}E_2 \rho(E_1)\rho(E_2) e^{-(\beta_{L1}+\beta_{R1})E_1-(\beta_{L2}+\beta_{R2})E_2}\langle E_1|\mathcal{O}_{q_1}\mathcal{O}_{-q_1}|E_2\rangle\langle E_2|\mathcal{O}_{q_2}\mathcal{O}_{-q_2}|E_1\rangle.
\end{split}
\ee

let us now look at the semi-classical geometry of the squared one-point function and two-point function using the asymptotic approximation of the gamma function:
\be
\Gamma(m+i x) \Gamma(m-i x)\sim 2\pi x^{2m-1} e^{-\pi x},~~~x\gg 1.
\ee
It is easy to see that both $|\langle \mathcal{O}_q\rangle|^2$ and $|\langle \mathcal{O}_{q_1}\mathcal{O}_{q_2}\rangle|^2$ are dominated by low-energy configurations:
\be
E\sim {m\over \beta_L+\beta_R}.
\ee
This energy is linear in temperature and the dimension of the inserted operator.
The effective temperature can be derived from the thermodynamical relation:
\be
\beta_E={\sqrt{2}\pi\over \sqrt{E}}.
\ee
With this information, we can look at the saddle configuration of the geodesic across the two asymptotic boundaries.
The integration of the geodesic length only contributes to the calculation of the two-point function in the energy basis:
\be
\begin{split}
\int \mathrm{d}\ell \langle \ell|E\rangle\langle E|\ell\rangle e^{-m\ell}=16\int \mathrm{d}\ell e^{-m\ell} K^2_{2i\sqrt{2E}}(4 e^{-\ell/2})\\
=4\int \mathrm{d}\ell \mathrm{d}t_1 \mathrm{d}t_2 e^{-m \ell-2i\sqrt{2E}(t_1+t_2)-4e^{-\ell/2}(\cosh t_1+\cosh t_2)}
\end{split}
\ee
where we use the integral representation of the Bessel K-function.
The saddle point equations are:
\be\label{eqn:saddle point}
4 e^{-\ell_*/ 2}\cosh t=m;~~~4 e^{-\ell_*/2}\sinh t=2i\sqrt{2E},
\ee
where $t$ is the saddle of $t_{1,2}$ and can be thought of as half of the (Lorentzian) Rindler angle spanned by the geodesic.
The solutions of the saddle point equations are:
\be
\ell_*=\log{16\over m^2+8 E},~~~t=i\arctan({2\sqrt{2E}\over m}).
\ee
Given $\ell_*,E$ and the length of the two asymptotic boundaries, the full geometry is specified. 
To fully characterize the geometry, it will be convenient to draw the wormhole geometry on the hyperbolic disk (figure \ref{fig:hyperdisk}) where we put the center of the two asymptotic boundaries at the center of the hyperbolic disk. In the Rindler coordinates, where the metric is $ds^2=d\rho^2+\sinh^2\rho d\theta^2$, this is at the location $\rho=0$.

Since the four images of the operators will all be at the same radius, we can denote their location as $(\rho_b,\theta_1)$ to $(\rho_b,\theta_4)$, where $\rho_b$ is purely determined by the energy $E$ and the IR cutoff $\epsilon$:
\be
\sinh\rho_b = {\beta_E\over 2\pi \epsilon}={1\over \sqrt{2E}\epsilon}.
\ee
The Rindler angle spanned by each of the asymptotic boundaries is also determined:
\be
\theta_{12,34}={2\pi \beta_{L,R}\over \beta_E}=\sqrt{2 E}\beta_{L,R}
\ee
The (regularized) geodesic distance between two bulk points $(\rho_b,\theta_2)$ and $(\rho_b,\theta_3)$ can be easily calculated by taking the inner product of their embedding coordinates, which gives:
\be
e^{ \ell}=\epsilon^2e^{2\rho_b} \sin^2({\theta_{23}\over 2})={2\over E}\sin^2({\theta_{23}\over 2})
\ee
Compared with the second equation for the saddle point \eqref{eqn:saddle point}, we confirm the statement that $t$ is half the Lorentzian Rindler angle spanned by the geodesic, $t=i{\theta_{23}\over 2}$.
An important quantity in this geometry is the length of the geodesic across the wormhole $b$ which characterizes the size of the wormhole.
Due to the SL(2,R) invariance, $b$ is a function of the cross ratio of the four $\theta$'s, and is therefore  fully determined:
\be\label{eqn:b-crossratio}
\sinh{b\over 2}=\sqrt{\sin{\theta_{12}\over 2}\sin{\theta_{34}\over 2}\over \sin{\theta_{32}\over 2}\sin{\theta_{42}\over 2}}={\sqrt{\sin{\pi\beta_L\over \beta_E}\sin{\pi\beta_R\over \beta_E}}\over |\sinh t|}.
\ee
Then equation \eqref{eqn:b-crossratio} tells us that the size of the wormhole grows linearly with time for $\beta_{L,R}\sim \pm i T$:
\be
b\sim {2\pi T\over \beta_E}+2\log(({m\beta_E\over 4\pi})^2+1)
\ee

 \section{State reconstruction in the canonical ensemble}
\label{sec:canonical-ensemble} 
 
In this appendix, we discuss the canonical ensemble version of the state reconstruction studied in section \ref{sec:state-reconst}.
In the case of the microcanonical ensemble, we see that the $K$ charge states span the whole Hilbert space once $K$ exceeds the number of states in the energy window, {\it i.e.} after the Page transition.

The canonical ensemble can be understood as a distribution of the microcanonical ensemble weighted by the Boltzmann factor, and we can separate the states into pre-Page states and post-Page states. This suggests that for fixed $K$, the charge states can span the subspace of all the post-Page states. As we increase $K$, this subspace will become larger as more and more states will hit the Page transition.  Once $K\gg e^{2S_\text{BH}}$, most of the states that dominate the thermal distribution will become post-Page and so any low-energy excitation will become reconstructable by the $K$ charge states.

We can make this picture more precise using the Schwinger-Dyson (SD) equation \eqref{eqn:SD equation}.
In the SD equation, we did not write down the explicit expression of $\rho(E)$ and $y(E_1,E_2)$.
In the limit of small $G_{\text N}$, they have the following approximation:
\be
\rho(E)\sim e^{2\pi \sqrt{2E}};~~~y(E_1,E_2)\sim e^{-S_0-2\pi \text{ max}(\sqrt{2E_1},\sqrt{2E_2})-{\beta\over 2}(E_1+E_2)}\,.
\ee
The precise form of these two functions is not that important; rather, we should just observe that $\rho(E)$ is a density of states that is increasing as a function of energy, and $y(E_1,E_2)$ is a combination of the Boltzmann factor with the matrix element of the operator and decreases with the energy.
This means that for fixed value of $R$, the integrand inside the SD equation \eqref{eqn:SD equation} can be approximated as two functions according to the energy when the energy is large:
\be
{y R\over 1-yR}\sim \begin{cases} -1;~~~~|y(E_1,E_2)R|>1;\\
y R;~~~~|y(E_1,E_2)R|<1.
\end{cases}
\ee
The contour that separates these two regions is determined by the equation $|y(E_1,E_2)R|=1$.
Roughly speaking, this is saying that states that have energy $|y(E_1,E_2)R|<1$ can be well approximated by just the cylinder geometry and states that have energy $|y(E_1,E_2)R>1|$ are well approximated by the pinwheel geometry.
Typically, $R$ has only one square root branch cut in the complex $\lambda$ plane and at infinite $\lambda$ it asymptotes to ${K\over \lambda}$.
This means that near the lower end $\lambda_0$ of the branch cut, $R$ is large and negative. 
So we can use the approximation to simplify the SD equation near $\lambda_0$:
\be\label{eqn:approximate_SDeqn}
\lambda\sim {K\over R}-{1\over R}\int_{-y(E_1,E_2)R>1} \mathrm{d}E_{1,2} \rho_{1,2}+\int_{-y(E_1,E_2)R<1} \mathrm{d}E_{1,2} \rho_{1,2} y,
\ee
where we use the notation $\mathrm{d}E_{1,2} \rho_{1,2}$ to represent the two energy integrals $\mathrm{d}E_1 \mathrm{d}E_2 e^{2S_0}\rho(E_1)\rho(E_2)$.
The branch point of the resolvent can therefore be determined by solving the equation ${d\lambda\over dR}=0$.
It is easy to see that the change of the integral domain cancels between the two integrals in equation \eqref{eqn:approximate_SDeqn}, and we are only left with equation:
\be
K=\int_{-y(E_1,E_2)R(\lambda_0)>1}\mathrm{d}E_{1,2}\rho_{1,2};~~~~\lambda_0=\int_{-y(E_1,E_2)R(\lambda_0)<1} \mathrm{d}E_{1,2} \rho_{1,2} y
\ee
The first equation determines the value of $R$ at $\lambda_0$.

The corresponding contour separates two regions in the energy plane, corresponding to the separation of post-Page states (states that have energy $-y(E_1,E_2)R(\lambda_0)>1$) and pre-Page states (states that have energy $-y(E_1,E_2)R(\lambda_0)<1$).
The second equation then tells us that all the pre-Page states contribute together as a shift of the end point of the branch cut, away from zero.
For $\lambda>\lambda_0$, the magnitude of $R$ will be smaller than $R(\lambda_0)$.
This means that the integral in the original SD equation \eqref{eqn:SD equation} can be separated into two regions based on whether the energies are pre-Page or post-Page:
\be\label{eqn:approxiSDequation}
\lambda R\sim K+\int_{-y(E_1,E_2)R(\lambda_0)>1} \mathrm{d}E_{1,2} \rho_{1,2}{y R\over 1-y R}+\lambda_0 R.
\ee
which in the limit of large $K$ and $\lambda$ away from $\lambda_0$, it has the approximate solution:
\be
R\sim {K\over \lambda-\lambda_0}.
\ee
Plugging this in the equation for $:VM^{-1}V:$ given by \eqref{eqn:VRV1} and deforming the integration contour of $\lambda$ around $\lambda_0$, we find:
\be
:VM^{-1} V:= \int \mathrm{d}E_{1,2}\rho_{1,2} y \oint {\mathrm{d}\lambda\over 2\pi i\lambda}{y R\over 1- yR}\sim \int \mathrm{d}E_{1,2}\rho_{1,2} y {yK\over \lambda_0+yK}\,.
\ee

Once again, this separate the energy integral into two regions depending on the relative value between $\lambda_0$ and $y K$:
for the energy range where $yK>\lambda_0$, we have a contribution $\int_{yK>\lambda_0} dE_{1,2}\rho_{1,2} y $
which reconstructs the energy component of state $|\psi\rangle$ in this region;
for the energy range where $yK<\lambda_0$, we have a contribution ${K\over \lambda_0}\int_{yK>\lambda_0} \mathrm{d}E_{1,2}\rho_{1,2} y^2 $ which quickly decays to zero when $K>e^{2S_\text{BH}}$.
As a consequence, this means that the fixed $K>e^{2S_\text{BH}}$ charge states span a complete basis in the states whose energy satisfies the following condition:
\be
y(E_1,E_2)>{\lambda_0\over K}.
\ee

\bibliographystyle{utphys}
\bibliography{main}{}

\providecommand{\href}[2]{#2}\begingroup\raggedright\begin{thebibliography}{10}

\bibitem{Misner:1957mt}
C.~W. Misner and J.~A. Wheeler, ``{Classical physics as geometry: Gravitation,
  electromagnetism, unquantized charge, and mass as properties of curved empty
  space},'' \href{http://dx.doi.org/10.1016/0003-4916(57)90049-0}{{\em Annals
  Phys.} {\bfseries 2} (1957) 525--603}.

\bibitem{Giddings:1988cx}
S.~B. Giddings and A.~Strominger, ``{Loss of Incoherence and Determination of
  Coupling Constants in Quantum Gravity},''
  \href{http://dx.doi.org/10.1016/0550-3213(88)90109-5}{{\em Nucl. Phys. B}
  {\bfseries 307} (1988) 854--866}.

\bibitem{Kallosh:1995hi}
R.~Kallosh, A.~D. Linde, D.~A. Linde, and L.~Susskind, ``{Gravity and global
  symmetries},'' \href{http://dx.doi.org/10.1103/PhysRevD.52.912}{{\em Phys.
  Rev. D} {\bfseries 52} (1995) 912--935},
  \href{http://arxiv.org/abs/hep-th/9502069}{{\ttfamily arXiv:hep-th/9502069}}.

\bibitem{Polchinski:2003bq}
J.~Polchinski, ``{Monopoles, duality, and string theory},''
  \href{http://dx.doi.org/10.1142/S0217751X0401866X}{{\em Int. J. Mod. Phys. A}
  {\bfseries 19S1} (2004) 145--156},
  \href{http://arxiv.org/abs/hep-th/0304042}{{\ttfamily arXiv:hep-th/0304042}}.

\bibitem{ArkaniHamed:2006dz}
N.~Arkani-Hamed, L.~Motl, A.~Nicolis, and C.~Vafa, ``{The String landscape,
  black holes and gravity as the weakest force},''
  \href{http://dx.doi.org/10.1088/1126-6708/2007/06/060}{{\em JHEP} {\bfseries
  06} (2007) 060}, \href{http://arxiv.org/abs/hep-th/0601001}{{\ttfamily
  arXiv:hep-th/0601001}}.

\bibitem{Banks:2010zn}
T.~Banks and N.~Seiberg, ``{Symmetries and Strings in Field Theory and
  Gravity},'' \href{http://dx.doi.org/10.1103/PhysRevD.83.084019}{{\em Phys.
  Rev. D} {\bfseries 83} (2011) 084019},
  \href{http://arxiv.org/abs/1011.5120}{{\ttfamily arXiv:1011.5120 [hep-th]}}.

\bibitem{Harlow:2018jwu}
D.~Harlow and H.~Ooguri, ``{Constraints on Symmetries from Holography},''
  \href{http://dx.doi.org/10.1103/PhysRevLett.122.191601}{{\em Phys. Rev.
  Lett.} {\bfseries 122} no.~19, (2019) 191601},
  \href{http://arxiv.org/abs/1810.05337}{{\ttfamily arXiv:1810.05337
  [hep-th]}}.

\bibitem{Harlow:2018tng}
D.~Harlow and H.~Ooguri, ``{Symmetries in quantum field theory and quantum
  gravity},'' \href{http://arxiv.org/abs/1810.05338}{{\ttfamily
  arXiv:1810.05338 [hep-th]}}.

\bibitem{Harlow:2020bee}
D.~Harlow and E.~Shaghoulian, ``{Global symmetry, Euclidean gravity, and the
  black hole information problem},''
  \href{http://arxiv.org/abs/2010.10539}{{\ttfamily arXiv:2010.10539
  [hep-th]}}.

\bibitem{HAWKING1974}
S.~W. HAWKING, ``Black hole explosions?,''
  \href{http://dx.doi.org/10.1038/248030a0}{{\em Nature} {\bfseries 248}
  no.~5443, (Mar, 1974) 30--31}. \url{https://doi.org/10.1038/248030a0}.

\bibitem{Hawking1975}
S.~W. Hawking, ``Particle creation by black holes,''
  \href{http://dx.doi.org/10.1007/BF02345020}{{\em Communications in
  Mathematical Physics} {\bfseries 43} no.~3, (Aug, 1975) 199--220}.
  \url{https://doi.org/10.1007/BF02345020}.

\bibitem{Hawking:1982dj}
S.~Hawking, ``{The Unpredictability of Quantum Gravity},''
  \href{http://dx.doi.org/10.1007/BF01206031}{{\em Commun. Math. Phys.}
  {\bfseries 87} (1982) 395--415}.

\bibitem{Almheiri:2020cfm}
A.~Almheiri, T.~Hartman, J.~Maldacena, E.~Shaghoulian, and A.~Tajdini, ``{The
  entropy of Hawking radiation},''
  \href{http://arxiv.org/abs/2006.06872}{{\ttfamily arXiv:2006.06872
  [hep-th]}}.

\bibitem{ABBOTT1989687}
L.~Abbott and M.~B. Wise, ``Wormholes and global symmetries,''
  \href{http://dx.doi.org/https://doi.org/10.1016/0550-3213(89)90503-8}{{\em
  Nuclear Physics B} {\bfseries 325} no.~3, (1989) 687 -- 704}.
  \url{http://www.sciencedirect.com/science/article/pii/0550321389905038}.

\bibitem{Penington:2019kki}
G.~Penington, S.~H. Shenker, D.~Stanford, and Z.~Yang, ``{Replica wormholes and
  the black hole interior},'' \href{http://arxiv.org/abs/1911.11977}{{\ttfamily
  arXiv:1911.11977 [hep-th]}}.

\bibitem{Almheiri:2019qdq}
A.~Almheiri, T.~Hartman, J.~Maldacena, E.~Shaghoulian, and A.~Tajdini,
  ``{Replica Wormholes and the Entropy of Hawking Radiation},''
  \href{http://dx.doi.org/10.1007/JHEP05(2020)013}{{\em JHEP} {\bfseries 05}
  (2020) 013}, \href{http://arxiv.org/abs/1911.12333}{{\ttfamily
  arXiv:1911.12333 [hep-th]}}.

\bibitem{Maldacena:2004rf}
J.~M. Maldacena and L.~Maoz, ``{Wormholes in AdS},''
  \href{http://dx.doi.org/10.1088/1126-6708/2004/02/053}{{\em JHEP} {\bfseries
  02} (2004) 053}, \href{http://arxiv.org/abs/hep-th/0401024}{{\ttfamily
  arXiv:hep-th/0401024}}.

\bibitem{Witten:1998qj}
E.~Witten, ``{Anti-de Sitter space and holography},''
  \href{http://dx.doi.org/10.4310/ATMP.1998.v2.n2.a2}{{\em Adv. Theor. Math.
  Phys.} {\bfseries 2} (1998) 253--291},
  \href{http://arxiv.org/abs/hep-th/9802150}{{\ttfamily arXiv:hep-th/9802150}}.

\bibitem{Sachdev:2015efa}
S.~Sachdev, ``{Bekenstein-Hawking Entropy and Strange Metals},''
  \href{http://dx.doi.org/10.1103/PhysRevX.5.041025}{{\em Phys. Rev. X}
  {\bfseries 5} no.~4, (2015) 041025},
  \href{http://arxiv.org/abs/1506.05111}{{\ttfamily arXiv:1506.05111
  [hep-th]}}.

\bibitem{kitaevTalks}
A.~Kitaev, ``{Talks given at the Fundamental Physics Prize Symposium and KITP
  seminars},''.

\bibitem{Maldacena:2016hyu}
J.~Maldacena and D.~Stanford, ``{Remarks on the Sachdev-Ye-Kitaev model},''
  \href{http://dx.doi.org/10.1103/PhysRevD.94.106002}{{\em Phys. Rev.}
  {\bfseries D94} no.~10, (2016) 106002},
\href{http://arxiv.org/abs/1604.07818}{{\ttfamily arXiv:1604.07818 [hep-th]}}.
%%CITATION = ARXIV:1604.07818;%%.

\bibitem{Kitaev:2017awl}
A.~Kitaev and S.~J. Suh, ``{The soft mode in the Sachdev-Ye-Kitaev model and
  its gravity dual},'' \href{http://dx.doi.org/10.1007/JHEP05(2018)183}{{\em
  JHEP} {\bfseries 05} (2018) 183},
  \href{http://arxiv.org/abs/1711.08467}{{\ttfamily arXiv:1711.08467
  [hep-th]}}.

\bibitem{Teitelboim:1983ux}
C.~Teitelboim, ``{Gravitation and Hamiltonian Structure in Two Space-Time
  Dimensions},''
\href{http://dx.doi.org/10.1016/0370-2693(83)90012-6}{{\em Phys. Lett.}
  {\bfseries B126} (1983) 41--45}.
%%CITATION = PHLTA,B126,41;%%.

\bibitem{Jackiw:1984je}
R.~Jackiw, ``{Lower Dimensional Gravity},''
\href{http://dx.doi.org/10.1016/0550-3213(85)90448-1}{{\em Nucl. Phys.}
  {\bfseries B252} (1985) 343--356}.
%%CITATION = NUPHA,B252,343;%%.

\bibitem{Chen:2020ojn}
Y.~Chen and H.~W. Lin, ``{Signatures of global symmetry violation in relative
  entropies and replica wormholes},''
  \href{http://arxiv.org/abs/2011.06005}{{\ttfamily arXiv:2011.06005
  [hep-th]}}.

\bibitem{Israel:1967wq}
W.~Israel, ``{Event horizons in static vacuum space-times},''
  \href{http://dx.doi.org/10.1103/PhysRev.164.1776}{{\em Phys. Rev.} {\bfseries
  164} (1967) 1776--1779}.

\bibitem{Israel:1967za}
W.~Israel, ``{Event horizons in static electrovac space-times},''
  \href{http://dx.doi.org/10.1007/BF01645859}{{\em Commun. Math. Phys.}
  {\bfseries 8} (1968) 245--260}.

\bibitem{Carter:1971zc}
B.~Carter, ``{Axisymmetric Black Hole Has Only Two Degrees of Freedom},''
  \href{http://dx.doi.org/10.1103/PhysRevLett.26.331}{{\em Phys. Rev. Lett.}
  {\bfseries 26} (1971) 331--333}.

\bibitem{Robinson:1975bv}
D.~Robinson, ``{Uniqueness of the Kerr black hole},''
  \href{http://dx.doi.org/10.1103/PhysRevLett.34.905}{{\em Phys. Rev. Lett.}
  {\bfseries 34} (1975) 905--906}.

\bibitem{Bousso:1999xy}
R.~Bousso, ``{A Covariant entropy conjecture},''
  \href{http://dx.doi.org/10.1088/1126-6708/1999/07/004}{{\em JHEP} {\bfseries
  07} (1999) 004}, \href{http://arxiv.org/abs/hep-th/9905177}{{\ttfamily
  arXiv:hep-th/9905177}}.

\bibitem{Bousso:2014sda}
R.~Bousso, H.~Casini, Z.~Fisher, and J.~Maldacena, ``{Proof of a Quantum Bousso
  Bound},'' \href{http://dx.doi.org/10.1103/PhysRevD.90.044002}{{\em Phys. Rev.
  D} {\bfseries 90} no.~4, (2014) 044002},
  \href{http://arxiv.org/abs/1404.5635}{{\ttfamily arXiv:1404.5635 [hep-th]}}.

\bibitem{Giddings:1993km}
S.~B. Giddings, ``{Constraints on black hole remnants},''
  \href{http://dx.doi.org/10.1103/PhysRevD.49.947}{{\em Phys. Rev. D}
  {\bfseries 49} (1994) 947--957},
  \href{http://arxiv.org/abs/hep-th/9304027}{{\ttfamily arXiv:hep-th/9304027}}.

\bibitem{Susskind:1995da}
L.~Susskind, ``{Trouble for remnants},''
  \href{http://arxiv.org/abs/hep-th/9501106}{{\ttfamily arXiv:hep-th/9501106}}.

\bibitem{Saad:2019pqd}
P.~Saad, ``{Late Time Correlation Functions, Baby Universes, and ETH in JT
  Gravity},'' \href{http://arxiv.org/abs/1910.10311}{{\ttfamily
  arXiv:1910.10311 [hep-th]}}.

\bibitem{Penington:2019npb}
G.~Penington, ``{Entanglement Wedge Reconstruction and the Information
  Paradox},'' \href{http://dx.doi.org/10.1007/JHEP09(2020)002}{{\em JHEP}
  {\bfseries 09} (2020) 002}, \href{http://arxiv.org/abs/1905.08255}{{\ttfamily
  arXiv:1905.08255 [hep-th]}}.

\bibitem{Almheiri:2019psf}
A.~Almheiri, N.~Engelhardt, D.~Marolf, and H.~Maxfield, ``{The entropy of bulk
  quantum fields and the entanglement wedge of an evaporating black hole},''
  \href{http://dx.doi.org/10.1007/JHEP12(2019)063}{{\em JHEP} {\bfseries 12}
  (2019) 063}, \href{http://arxiv.org/abs/1905.08762}{{\ttfamily
  arXiv:1905.08762 [hep-th]}}.

\bibitem{Almheiri:2019hni}
A.~Almheiri, R.~Mahajan, J.~Maldacena, and Y.~Zhao, ``{The Page curve of
  Hawking radiation from semiclassical geometry},''
  \href{http://dx.doi.org/10.1007/JHEP03(2020)149}{{\em JHEP} {\bfseries 03}
  (2020) 149}, \href{http://arxiv.org/abs/1908.10996}{{\ttfamily
  arXiv:1908.10996 [hep-th]}}.

\bibitem{Engelhardt:2014gca}
N.~Engelhardt and A.~C. Wall, ``{Quantum Extremal Surfaces: Holographic
  Entanglement Entropy beyond the Classical Regime},''
  \href{http://dx.doi.org/10.1007/JHEP01(2015)073}{{\em JHEP} {\bfseries 01}
  (2015) 073}, \href{http://arxiv.org/abs/1408.3203}{{\ttfamily arXiv:1408.3203
  [hep-th]}}.

\bibitem{PhysRevD.13.2188}
J.~B. Hartle and S.~W. Hawking, ``Path-integral derivation of black-hole
  radiance,'' \href{http://dx.doi.org/10.1103/PhysRevD.13.2188}{{\em Phys. Rev.
  D} {\bfseries 13} (Apr, 1976) 2188--2203}.
  \url{https://link.aps.org/doi/10.1103/PhysRevD.13.2188}.

\bibitem{PhysRevD.28.2960}
J.~B. Hartle and S.~W. Hawking, ``Wave function of the universe,''
  \href{http://dx.doi.org/10.1103/PhysRevD.28.2960}{{\em Phys. Rev. D}
  {\bfseries 28} (Dec, 1983) 2960--2975}.
  \url{https://link.aps.org/doi/10.1103/PhysRevD.28.2960}.

\bibitem{Maldacena:2001ss}
J.~M. Maldacena, G.~W. Moore, and N.~Seiberg, ``{D-brane charges in five-brane
  backgrounds},'' \href{http://dx.doi.org/10.1088/1126-6708/2001/10/005}{{\em
  JHEP} {\bfseries 10} (2001) 005},
  \href{http://arxiv.org/abs/hep-th/0108152}{{\ttfamily arXiv:hep-th/0108152}}.

\bibitem{Kapustin:2014gua}
A.~Kapustin and N.~Seiberg, ``{Coupling a QFT to a TQFT and Duality},''
  \href{http://dx.doi.org/10.1007/JHEP04(2014)001}{{\em JHEP} {\bfseries 04}
  (2014) 001}, \href{http://arxiv.org/abs/1401.0740}{{\ttfamily arXiv:1401.0740
  [hep-th]}}.

\bibitem{Saad:2019lba}
P.~Saad, S.~H. Shenker, and D.~Stanford, ``{JT gravity as a matrix integral},''
  \href{http://arxiv.org/abs/1903.11115}{{\ttfamily arXiv:1903.11115
  [hep-th]}}.

\bibitem{Stanford:2020wkf}
D.~Stanford, ``{More quantum noise from wormholes},''
  \href{http://arxiv.org/abs/2008.08570}{{\ttfamily arXiv:2008.08570
  [hep-th]}}.

\bibitem{Mertens:2017mtv}
T.~G. Mertens, G.~J. Turiaci, and H.~L. Verlinde, ``{Solving the Schwarzian via
  the Conformal Bootstrap},''
  \href{http://dx.doi.org/10.1007/JHEP08(2017)136}{{\em JHEP} {\bfseries 08}
  (2017) 136}, \href{http://arxiv.org/abs/1705.08408}{{\ttfamily
  arXiv:1705.08408 [hep-th]}}.

\bibitem{Kitaev:2018wpr}
A.~Kitaev and S.~J. Suh, ``{Statistical mechanics of a two-dimensional black
  hole},'' \href{http://dx.doi.org/10.1007/JHEP05(2019)198}{{\em JHEP}
  {\bfseries 05} (2019) 198}, \href{http://arxiv.org/abs/1808.07032}{{\ttfamily
  arXiv:1808.07032 [hep-th]}}.

\bibitem{Yang:2018gdb}
Z.~Yang, ``{The Quantum Gravity Dynamics of Near Extremal Black Holes},''
  \href{http://dx.doi.org/10.1007/JHEP05(2019)205}{{\em JHEP} {\bfseries 05}
  (2019) 205}, \href{http://arxiv.org/abs/1809.08647}{{\ttfamily
  arXiv:1809.08647 [hep-th]}}.

\bibitem{Blommaert:2018oro}
A.~Blommaert, T.~G. Mertens, and H.~Verschelde, ``{The Schwarzian Theory - A
  Wilson Line Perspective},''
  \href{http://dx.doi.org/10.1007/JHEP12(2018)022}{{\em JHEP} {\bfseries 12}
  (2018) 022}, \href{http://arxiv.org/abs/1806.07765}{{\ttfamily
  arXiv:1806.07765 [hep-th]}}.

\bibitem{Iliesiu:2019xuh}
L.~V. Iliesiu, S.~S. Pufu, H.~Verlinde, and Y.~Wang, ``{An exact quantization
  of Jackiw-Teitelboim gravity},''
  \href{http://dx.doi.org/10.1007/JHEP11(2019)091}{{\em JHEP} {\bfseries 11}
  (2019) 091}, \href{http://arxiv.org/abs/1905.02726}{{\ttfamily
  arXiv:1905.02726 [hep-th]}}.

\bibitem{Witten:1991we}
E.~Witten, ``{On quantum gauge theories in two-dimensions},''
  \href{http://dx.doi.org/10.1007/BF02100009}{{\em Commun. Math. Phys.}
  {\bfseries 141} (1991) 153--209}.

\bibitem{Witten:1992xu}
E.~Witten, ``{Two-dimensional gauge theories revisited},''
  \href{http://dx.doi.org/10.1016/0393-0440(92)90034-X}{{\em J. Geom. Phys.}
  {\bfseries 9} (1992) 303--368},
  \href{http://arxiv.org/abs/hep-th/9204083}{{\ttfamily arXiv:hep-th/9204083}}.

\bibitem{Iliesiu:2019lfc}
L.~V. Iliesiu, ``{On 2D gauge theories in Jackiw-Teitelboim gravity},''
  \href{http://arxiv.org/abs/1909.05253}{{\ttfamily arXiv:1909.05253
  [hep-th]}}.

\bibitem{Kapec:2019ecr}
D.~Kapec, R.~Mahajan, and D.~Stanford, ``{Matrix ensembles with global
  symmetries and \textquoteright{}t Hooft anomalies from 2d gauge theory},''
  \href{http://dx.doi.org/10.1007/JHEP04(2020)186}{{\em JHEP} {\bfseries 04}
  (2020) 186}, \href{http://arxiv.org/abs/1912.12285}{{\ttfamily
  arXiv:1912.12285 [hep-th]}}.

\bibitem{Maxfield:2020ale}
H.~Maxfield and G.~J. Turiaci, ``{The path integral of 3D gravity near
  extremality; or, JT gravity with defects as a matrix integral},''
  \href{http://arxiv.org/abs/2006.11317}{{\ttfamily arXiv:2006.11317
  [hep-th]}}.

\bibitem{Witten:2020wvy}
E.~Witten, ``{Matrix Models and Deformations of JT Gravity},''
  \href{http://arxiv.org/abs/2006.13414}{{\ttfamily arXiv:2006.13414
  [hep-th]}}.

\bibitem{Gaiotto:2014kfa}
D.~Gaiotto, A.~Kapustin, N.~Seiberg, and B.~Willett, ``{Generalized Global
  Symmetries},'' \href{http://dx.doi.org/10.1007/JHEP02(2015)172}{{\em JHEP}
  {\bfseries 02} (2015) 172}, \href{http://arxiv.org/abs/1412.5148}{{\ttfamily
  arXiv:1412.5148 [hep-th]}}.

\bibitem{Stanford:2019vob}
D.~Stanford and E.~Witten, ``{JT Gravity and the Ensembles of Random Matrix
  Theory},'' \href{http://arxiv.org/abs/1907.03363}{{\ttfamily arXiv:1907.03363
  [hep-th]}}.

\bibitem{Marolf:2020xie}
D.~Marolf and H.~Maxfield, ``{Transcending the ensemble: baby universes,
  spacetime wormholes, and the order and disorder of black hole information},''
  \href{http://dx.doi.org/10.1007/JHEP08(2020)044}{{\em JHEP} {\bfseries 08}
  (2020) 044}, \href{http://arxiv.org/abs/2002.08950}{{\ttfamily
  arXiv:2002.08950 [hep-th]}}.

\bibitem{Dong:2020iod}
X.~Dong and H.~Wang, ``{Enhanced corrections near holographic entanglement
  transitions: a chaotic case study},''
  \href{http://dx.doi.org/10.1007/JHEP11(2020)007}{{\em JHEP} {\bfseries 11}
  (2020) 007}, \href{http://arxiv.org/abs/2006.10051}{{\ttfamily
  arXiv:2006.10051 [hep-th]}}.

\bibitem{Marolf:2020vsi}
D.~Marolf, S.~Wang, and Z.~Wang, ``{Probing phase transitions of holographic
  entanglement entropy with fixed area states},''
  \href{http://arxiv.org/abs/2006.10089}{{\ttfamily arXiv:2006.10089
  [hep-th]}}.

\bibitem{Akers:2020pmf}
C.~Akers and G.~Penington, ``{Leading order corrections to the quantum extremal
  surface prescription},'' \href{http://arxiv.org/abs/2008.03319}{{\ttfamily
  arXiv:2008.03319 [hep-th]}}.

\bibitem{Iliesiu:2020qvm}
L.~V. Iliesiu and G.~J. Turiaci, ``{The statistical mechanics of near-extremal
  black holes},'' \href{http://arxiv.org/abs/2003.02860}{{\ttfamily
  arXiv:2003.02860 [hep-th]}}.

\bibitem{Nayak:2018qej}
P.~Nayak, A.~Shukla, R.~M. Soni, S.~P. Trivedi, and V.~Vishal, ``{On the
  Dynamics of Near-Extremal Black Holes},''
  \href{http://dx.doi.org/10.1007/JHEP09(2018)048}{{\em JHEP} {\bfseries 09}
  (2018) 048},
\href{http://arxiv.org/abs/1802.09547}{{\ttfamily arXiv:1802.09547 [hep-th]}}.
%%CITATION = ARXIV:1802.09547;%%.

\bibitem{Moitra:2018jqs}
U.~Moitra, S.~P. Trivedi, and V.~Vishal, ``{Extremal and near-extremal black
  holes and near-CFT$_{1}$},''
  \href{http://dx.doi.org/10.1007/JHEP07(2019)055}{{\em JHEP} {\bfseries 07}
  (2019) 055},
\href{http://arxiv.org/abs/1808.08239}{{\ttfamily arXiv:1808.08239 [hep-th]}}.
%%CITATION = ARXIV:1808.08239;%%.

\end{thebibliography}\endgroup

\end{document}